\documentclass[useAMS,usenatbib,letterpaper]{mn2e}

\pdfoutput=1
\usepackage{psfig} 
\usepackage{amssymb} 
\usepackage[pdftex]{graphicx}

\def\btg{G\"{a}nsicke}
\def\ecs{ergs~cm$^{-2}$~s$^{-1}$}
\def\ha{H$\alpha$}
\def\msun{$M_\odot\,$}
\def\Porb{P_{\mbox{\tiny orb}}}
\def\Pmin{P_{\mbox{\tiny min}}}
\def\Kem{K_{\mbox{\tiny em}}}
\def\rosat{{\em ROSAT}}
\def\galex{{\em GALEX}}
\def\rsun{$R_\odot\,$}
\def\Twd{$T_{\mbox{\tiny wd}}$}
\def\Teff{$T_{\mbox{\tiny eff}}$}
\newcommand{\be}{\begin{equation}}
\newcommand{\ee}{\end{equation}}

\title[Magnetic white dwarf binaries]{The evolutionary state of short period magnetic white dwarf binaries}
\author[E. Breedt et al.]{E.~Breedt$^{1}$\thanks{E-mail: e.breedt@warwick.ac.uk},
B.T.~\btg$^{1}$, J.~Girven$^{1}$, A.J.~Drake$^{2}$, C.M.~Copperwheat$^{1}$, \and S.G.~Parsons$^{1}$, T.R.~Marsh$^{1}$\\
$^{1}$ Department of Physics, University of Warwick, Coventry, UK\\
$^{2}$ California Institute of Technology, 1200 E. California Blvd, CA 91225, USA}

\begin{document}

\date{Accepted/Received}

\pagerange{\pageref{firstpage}--\pageref{lastpage}} \pubyear{2012}

\maketitle

\begin{abstract}
We present phase-resolved spectroscopy of two new short period 
low accretion rate magnetic binaries, SDSSJ125044.42+154957.3 ($\Porb=86$~min) and
SDSSJ151415.65+074446.5 ($\Porb=89$~min). Both systems were previously identified as magnetic
white dwarfs from the Zeeman splitting of the Balmer absorption lines in
their optical spectra. Their spectral energy distributions exhibit a large 
near-infrared excess, which we interpret as a combination of cyclotron emission 
and possibly a late type companion star. No absorption features from the companion are 
seen in our optical spectra. We derive the orbital periods from a narrow,
variable \ha\, emission line which we show to originate on the
companion star. The high radial velocity amplitude measured in both systems suggests a
high orbital inclination, but we find no evidence for eclipses in our data. The two new systems
resemble the polar EF~Eri in its prolonged low state and also 
SDSSJ121209.31+013627.7, a known magnetic white dwarf plus possible brown dwarf binary, 
which was also recovered by our method. 
\end{abstract}



%

\section{Introduction}     \label{sec:intro}

Magnetic fields are commonly observed in white dwarfs. Field strengths of 1~MG or greater are found in at least $10$ per cent of isolated white dwarfs \citep{liebert03}, but their formation mechanism is not yet understood. 

It is generally thought that the magnetic fields in white dwarfs are `fossil fields', i.e. a remnant product of the evolution of chemically peculiar stars of spectral type A and B \citep{moss87,braithwaitespruit04}. Although a plausible formation model, the current space density of Ap and Bp stars cannot account for all the magnetic white dwarfs observed \citep{kawkavennes04,wickramasingheferrario05}. If the
progenitors of magnetic white dwarfs are single stars, as suggested by this model, we might expect the same fraction of magnetic white dwarfs among binary systems as found among single stars. 
Observations, however, show quite the opposite. Among the 2248 detached white dwarf plus main sequence binary systems found in the Sloan Digital Sky Survey (SDSS), not one was found to host a magnetic white dwarf \citep[][but see also \citealt{liebert05}, where this problem was first addressed]{rebassa11}. 

Semi-detached magnetic white dwarf binaries, on the other hand, are commonly found. They are known as polars or intermediate polars, based on the strength of the white dwarf magnetic field. They are a subclass of the cataclysmic variable stars (CVs), in which mass transfer occurs from an M-dwarf companion star to a white dwarf via Roche lobe overflow. In polars, the magnetic field of the white dwarf is strong enough ($B\gtrsim10$~MG) that it channels the accretion flow directly onto the white dwarf and no accretion disc is formed. 
Magnetic CVs account for $\sim25$ per cent of the 926 known CVs \citep[][v.7.16, 2011]{rkcat} --- a considerably higher fraction than is observed among isolated white dwarfs.

This apparent propensity for magnetic white dwarfs to form as part of interacting binary systems, has led \citet{tout08} to develop an alternative model for their formation, proposing that all high field magnetic white dwarfs have a binary origin. The model is based on the standard common envelope (CE) binary evolution theory of \citet{paczynski76}. When the more massive star in the binary evolves through the red giant phase, it expands and envelops both stars. The stars continue spiralling around one another inside the CE. According to their model, strong dynamo fields are generated by the orbiting stars in the core of the CE \citep[see also][]{pottertout10}. As the stars lose angular momentum and spiral closer together into the centre of the CE, the magnetic field is frozen into the degenerate stellar core. The product of the CE evolution depends on the final separation of the components when the envelope is expelled. Binaries with a wide separation will have no or very little frozen-in magnetic field on the white dwarf, and so leave the CE as non-magnetic pre-CVs or non-interacting non-magnetic wide binaries. For a strong magnetic field to form in the white dwarf, the binary must leave the CE with a small orbital separation. As a result, all magnetic binaries are close enough to be interacting, either through Roche lobe overflow, or by wind accretion. In the case where the two stars merge inside the CE, it will form a single, massive magnetic white dwarf \citep[see also][]{nordhaus11}. 

Wind-accreting magnetic binaries, as predicted by this model, are a recent discovery. The first such system, WX~LMi \citep[originally known as HS1023+3900,][]{reimers99_hs1023}, was discovered in the Hamburg Quasar Survey \citep[HQS,][]{hagen95} during follow-up spectroscopy of quasar candidates. A further eight systems have been discovered since (see \citealt{schmidt07} for a summary). All are faint or undetected in X-rays, have cool white dwarfs (\Twd$\lesssim10\,000$~K), and display strong cyclotron emission features in their optical or near-infrared (NIR) spectra. The accretion rates inferred from these characteristics are two to three orders of magnitude lower than typically observed in polars, so initially they were referred to as ``low accretion rate polars'' \citep{schwope02larps}. However, it was later realised that the secondary stars in these systems are underfilling their Roche lobes \citep[e.g.][]{schmidt05b}, distinguishing them from true polars. The inferred low accretion rates are consistent with magnetic siphoning of the secondary's stellar wind \citep{ww05}. These low accretion rate systems are now recognised as pre-polars, i.e. the progenitors of magnetic CVs, and they are expected to start Roche lobe overflow in the future as they evolve
to shorter periods. The nine systems discovered so far lead to an estimated space density of $\sim10^{-7}$~pc$^{-3}$ \citep{ww05}, more than a factor of ten lower than that of polars \citep{araujobetancor05b,thomasbeuermann98}.

Identifying a newly discovered system as a pre-polar is not straightforward, as they share many observable properties with low state polars. These ``low states'' are a well-known observed property of polars during which the Roche lobe overflow is temporarily interrupted and the accretion rate drops by an order of magnitude or more. These low accretion states are irregular and unpredictable, and can last for days to 
months \citep{kafka09amher}. During this time, polars appear essentially detached, so it is not immediately obvious to which class a newly discovered system belongs. For example, EQ~Cet was discovered in a low state \citep{schwope99} and thought to be a pre-polar, but it was later detected in a high state \citep{schwope02rosat} and dropped from the class.

The reason for the interrupted accretion flow is not yet fully understood. Starspots on the secondary star moving past the L1 point has been suggested as a possible cause of this behaviour \citep{liviopringle94}, but it seems an unlikely explanation for the extended low states seen in some polars, which can persist for decades. EF~Eri is the best-known example of such a low state polar. It was known as a conventional polar for almost two decades, before it dropped into a low state in 1997 \citep{wheatley_eferilow}. Apart from short bursts of accretion \citep{howell06}, it has not fully returned to a high state since. With an orbital period of 81~minutes, it is close to the orbital period minimum for CVs \citep{gaensicke09}, and a good candidate for having a substellar companion \citep{beuermann00,schwope10_eferixsh}. The companion star has not been directly detected yet, and a stellar companion cannot be ruled out completely \citep{howell06}.  

Evidence for a substellar companion to a cool magnetic white dwarf is also seen in SDSSJ121209.31+013627.7 (hereafter SDSSJ1212+0136). It was first identified as a binary by \citet{schmidt05b}, from its optical spectrum. In addition to the hallmarks of a magnetic white dwarf, the spectrum also showed a narrow \ha\, emission line, variable both in strength and radial velocity on a period of $\sim90$~minutes. \citet{schmidt05b} favoured a pre-polar interpretation of the system, but \citet{debes06} showed that the $K$ band variability was due to cyclotron emission, suggesting that there is ongoing low-level accretion in this system. Variability in the optical is pulse-like and consistent with emission from a self-eclipsing heated 
polar cap, rotating into and out of view every 88~minutes \citep{koenmaxted06,burleigh06,linnell10}. This type of optical variability is commonly seen in low state polars, including EF~Eri \citep[e.g.][]{ferrario92v834cen,howell06}.  \citet{burleigh06} also detected X-ray emission from this heated polar cap, and were able to constrain the accretion rate in SDSSJ1212+0136 to $\dot{M}\sim10^{-13}$\msun yr$^{-1}$. 
Cyclotron emission dominates the NIR part of the spectrum and complicates a direct spectroscopic detection of the companion star, but \citet{farihi08} showed that this part of the spectrum can be described by a combination of cyclotron emission and an L8 brown dwarf companion star. 

In this paper we present time-resolved optical spectroscopy of two more such short period magnetic binary systems, SDSSJ125044.42+154957.3 and SDSSJ151415.65+074446.5 (hereafter SDSSJ1250+1549 and SDSSJ1514+0744). 
Both systems have previously been identified as magnetic white dwarfs from their SDSS spectra \citep{vanlandingham05,kulebi09} and SDSSJ1250+1549 was noted by \citet{steele11} as having an infrared excess,
possibly due to an unresolved companion. The observation that SDSSJ1514+0744 is a binary system is a new discovery. In the following sections, we derive constraints on the binary parameters of these two systems and discuss their evolutionary state as polars or pre-polars in comparison to similar systems.


%
\section{Discovery and follow-up observations}

\subsection{SDSS and UKIDSS cross-matching} \label{sec:crossmatch}

The two new magnetic binaries were discovered as part of a larger project of identifying hydrogen dominated (DA) white dwarfs from their colours, rather than through spectroscopy. \citet{girven11} applied colour cuts to the SDSS Data Release 7 (DR7) photometry to select DA white dwarfs, and then fit the $ugri$ magnitudes with a grid of non-magnetic white dwarf models to determine the white dwarf parameters.  SDSS spectroscopy is available for 44\% of the sources which were identified as DA white dwarfs in this way, and served as an robust test of the efficiency of the selection method.
The resulting catalogue of objects was then cross-correlated with the Large Area Survey (LAS) of the United Kingdom InfraRed Telescope (UKIRT) Infrared Deep Sky Survey (UKIDSS) Data Release 8. This allowed us to identify white dwarfs with low mass stellar companions, substellar companions or debris discs, through a near-infrared flux excess over the extrapolated best-fit white dwarf model.

\begin{figure}
\centering
\rotatebox{270}{\includegraphics[width=7.0cm]{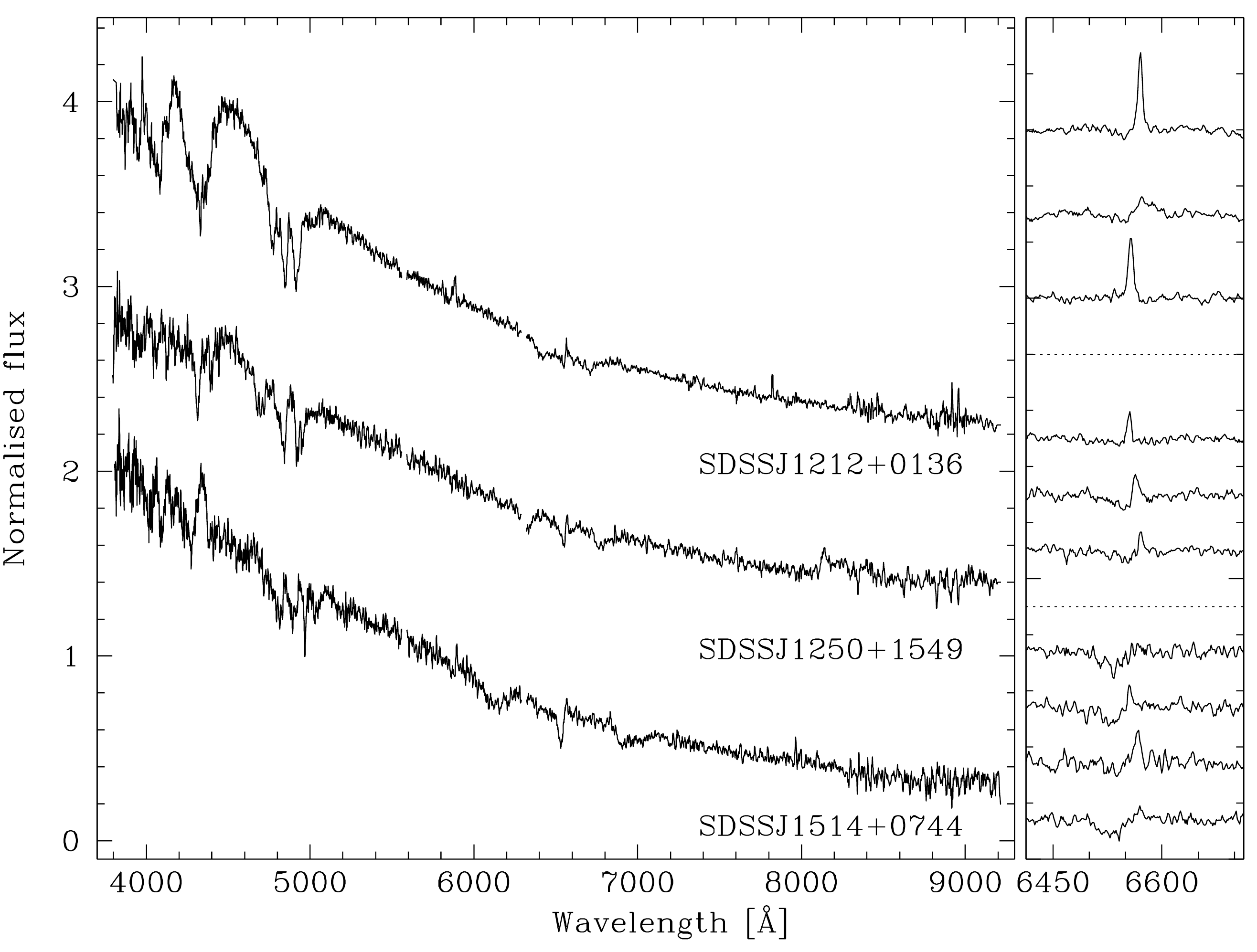}}
\caption{\label{fig:sdssspectra} {\em Main panel:} The SDSS spectra of  SDSSJ1212+0136,
SDSSJ1250+1549 and SDSSJ1514+0744, showing Zeeman splitting of the hydrogen absorption 
lines in dipolar fields of 13, 20 and 36 MG respectively. All three show weak \ha\, 
emission but no evidence of a companion star.
{\em Small panels:} SDSS subspectra of the three targets, in the same vertical order. 
The spectra are centred on \ha\, and are offset for clarity.
Variation in radial velocity and strength of the \ha\, line is clearly visible.}
\end{figure}

Full details of the cross-matching, selection criteria, and the efficiency of the method may be found in \citet{girven11}, but we note here that 51 of the 4636 spectroscopically confirmed white dwarfs were found to have an infrared excess. Three of these are of interest in this paper: SDSSJ1212+0136, SDSSJ1250+1549 and 
SDSSJ1514+0744. These three white dwarfs are clearly magnetic, and were previously identified as such by \citet{schmidt03}, \citet{vanlandingham05} and \citet{kulebi09}, respectively. The SDSS spectra are shown in Figure~\ref{fig:sdssspectra}; Zeeman splitting of the Balmer absorption lines is clearly visible. The dipolar field strengths derived from these spectra in their discovery papers are 13, 20 and 36~MG respectively.

\begin{figure}
\centering
\includegraphics[width=8.5cm]{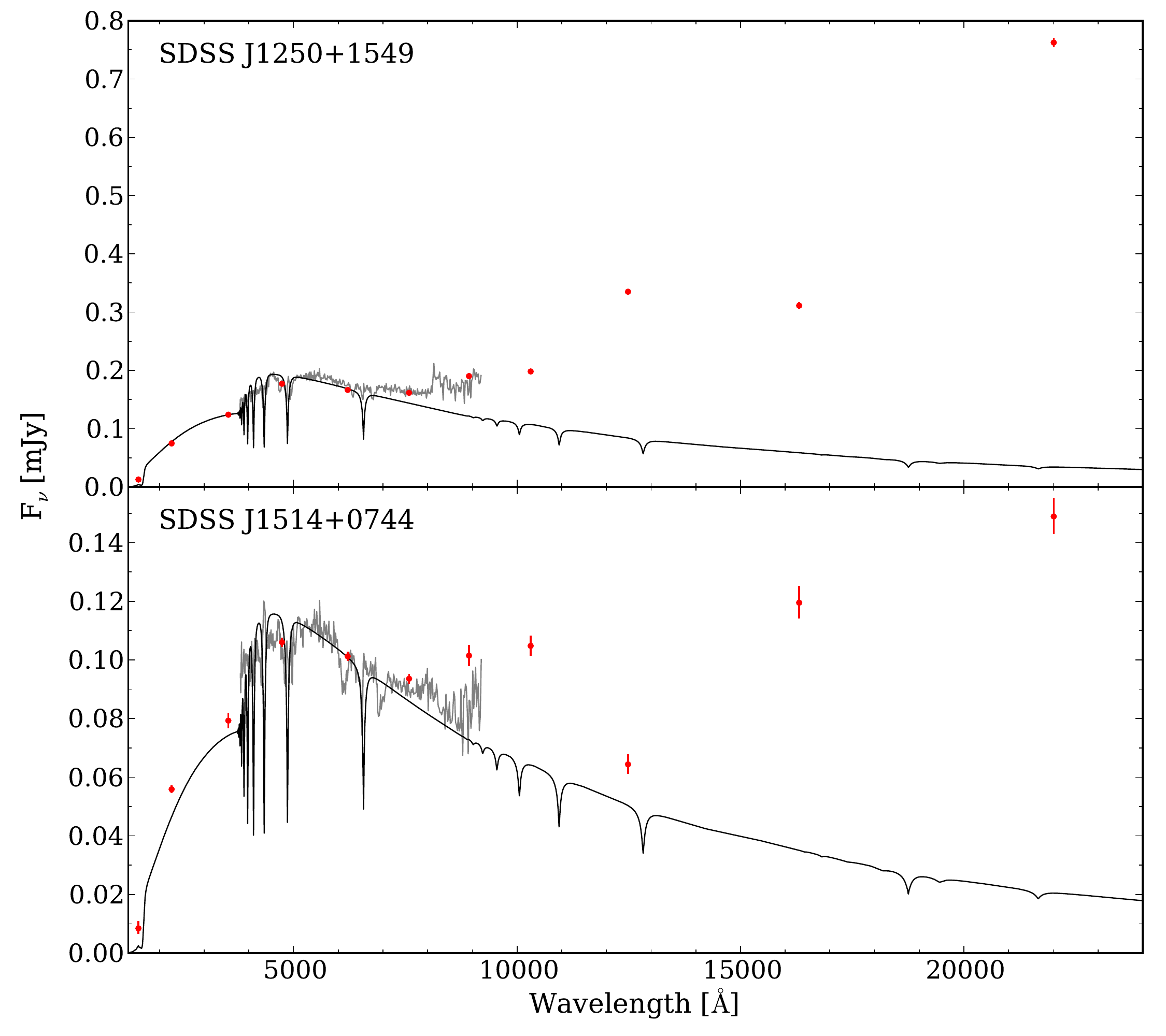}
\caption{\label{fig:seds} The Spectral Energy Distribution of SDSSJ1250+1549 (top) and 
SDSSJ1514+0744 (bottom). The measured \galex, SDSS $ugriz$ and UKIDSS $YJHK$ photometry
are plotted as red dots, the SDSS spectrum as a grey line and a 10\,000~K non-magnetic 
DA white dwarf model as a solid black line. The white dwarf temperature was derived from
a fit to the SDSS $ugri$ photometry (see \citealt{girven11} for details).
These systems were identified as binaries during a search for DA white dwarfs in the 
SDSS, when the cross-match with UKIDSS objects revealed their strong infrared excesses. 
}
\end{figure}

As discussed in Section~\ref{sec:intro}, SDSSJ1212+0136 is a known low accretion rate magnetic white dwarf plus brown dwarf binary. The two new binaries, SDSSJ1250+1549 and SDSSJ1514+0744, are the main subject of this paper. Figure~\ref{fig:seds} shows their spectral energy distributions, as derived from \galex\, ({\em Galaxy Evolution Explorer}), SDSS, and UKIDSS observations. A model spectrum of a 10\,000~K non-magnetic DA white dwarf is overplotted on the photometric measurements, highlighting the strong infrared excess observed in both systems.

\subsection{Time-resolved spectroscopy}

SDSS spectra are made up of several subexposures, which are individually extracted and combined to produce a final spectrum with higher signal to noise. For SDSSJ1250+1549 there are three subspectra available, taken over 56 minutes. For SDSSJ1514+0744 there are four subspectra, taken over 101 minutes, and another three spectra, of poorer quality, taken on a previous day. Inspection of these subspectra showed clear radial velocity variations of the \ha\, emission line in SDSSJ1250+1549 and a weak indication of a moving \ha\,
line in SDSSJ1514+0744 (Figure~\ref{fig:sdssspectra}). These results suggested that the two new targets may be binaries as well, and prompted us to carry out time-resolved spectroscopy on them for a more detailed investigation.

We observed SDSSJ1250+1549 and SDSSJ1514+0744 on the nights of 2011 March 24 -- 26 using the FORS2 spectrograph mounted on Unit Telescope 1 of the Very Large Telescope (VLT) at Paranal, Chile. Conditions were clear with subarcsecond seeing on all three nights. We used the GRIS1200R+76 grism with a 1\arcsec\, 
slit and standard $2\times2$ binning, which provided spectral coverage of the wavelength range 5872 -- 7370\AA\, dispersed at 0.73\AA\, per binned pixel. The resolution was estimated to be 1.7\AA,\, based on the measured full width at half maximum (FWHM) of strong skylines. 

We obtained 24 spectra of SDSSJ1250+1549 on the first night, using an exposure time of 360~s, a further six spectra on the second night and two on the third night. We used a 540~s exposure for SDSSJ1514+0744, obtaining six spectra on the first night, 13 on the second and 30 on the third. 

The data reduction and spectral extraction were carried out using {\sc pamela} \citep{marsh89pamela}, which is now available as part of the latest {\sc starlink}\footnote{http://starlink.jach.hawaii.edu/starlink. {\sc pamela} has been included since the `Hawaiki' release.} distribution. The wavelength scale was derived from a fifth order polynomial fit to 25 arclines using {\sc molly}\footnote{{\sc molly} was written by TRM and is available from http://www.warwick.ac.uk/go/trmarsh/software/.}. The root mean square (RMS) of the fit residuals was 0.012\AA.\, The helium-argon-neon arc lamp exposures were taken during the day, so to account for small shifts due to flexure of the spectrograph, we shifted the individual spectra by the amount required to move strong skylines to their known wavelengths. The shifts were typically $0.1-0.2$ pixels.

The spectra of the two objects display similar characteristics: large radial velocity variations, variable \ha\, emission and Zeeman splitting of the white dwarf \ha\, absorption line. No features of a companion star are visible in our spectra.

\begin{table*} 
  \centering
  \caption{\label{tab:mag} \galex, SDSS $ugriz$ and UKIDSS $YJHK$ magnitudes of the new
magnetic binaries.}
  \begin{tabular}{lccccccccccc} \hline
     SDSS name                & $FUV$ & $NUV$ &  $u$  &  $g$  &  $r$  &  $i$  & $z$  &  $Y$  &  $J$  &  $H$  & $K$    \\
     \hline
     SDSS J125044.42+154957.3 & 21.15 & 19.22 & 18.65 & 18.27 & 18.33 & 18.36 & 18.20 & 17.53 & 16.65 & 16.29 & 14.80 \\
     SDSS J151415.65+074446.5 & 21.64 & 19.39 & 19.16 & 18.84 & 18.88 & 18.99 & 18.88 & 18.22 & 18.44 & 17.33 & 16.57 \\
     \hline 
  \end{tabular}
\end{table*}

\subsection{\galex\, ultraviolet fluxes}

Both targets are listed in the \galex\, all sky survey catalogue, with strong near-ultraviolet ($NUV$, $\lambda_{\mbox{\small eff}}\sim2267$\AA) and slightly weaker far-ultraviolet ($FUV$, $\lambda_{\mbox{\small eff}}\sim1516$\AA) fluxes. The ultraviolet magnitudes from \galex\, Data Release 6 (GR6) are shown in Table~\ref{tab:mag} and Figure~\ref{fig:seds}.

\subsection{\rosat\, non-detection}

SDSSJ1250+1549 falls in the field of view of three \rosat\, ({\em R\"{o}ntgensatellit}) Position Sensitive Proportional Counter (PSPC) pointings, the longest of which had an effective exposure time of 8857 seconds
at the position of our target (dataset ID RP800393A01, 1992 December 17). It is not detected in any of the images. There were also two PSPC pointings which included SDSSJ1514+0744 in the field of view, the longest lasting for 1182 
seconds, corrected for vignetting effects (dataset ID RP800275N00, 1992 August 12). It is also undetected.  
These non-detections limit the X-ray count rates to $< 10^{-2}$~counts~s$^{-1}$ at the time of the \rosat\, 
observations.

\subsection{Optical photometry}

We carried out simple differential photometry (using {\sc iraf}) on our $R$ band acquisition images, and converted the magnitudes to the SDSS-$r$ band using the transformation equations of \citet{jester05sdss}. The magnitudes were
constant within photometric error over the three nights of observations and also consistent with the SDSS DR7 magnitudes as listed in Table~\ref{tab:mag}. 

\begin{figure}
\centering
\rotatebox{0}{\includegraphics[width=9.0cm]{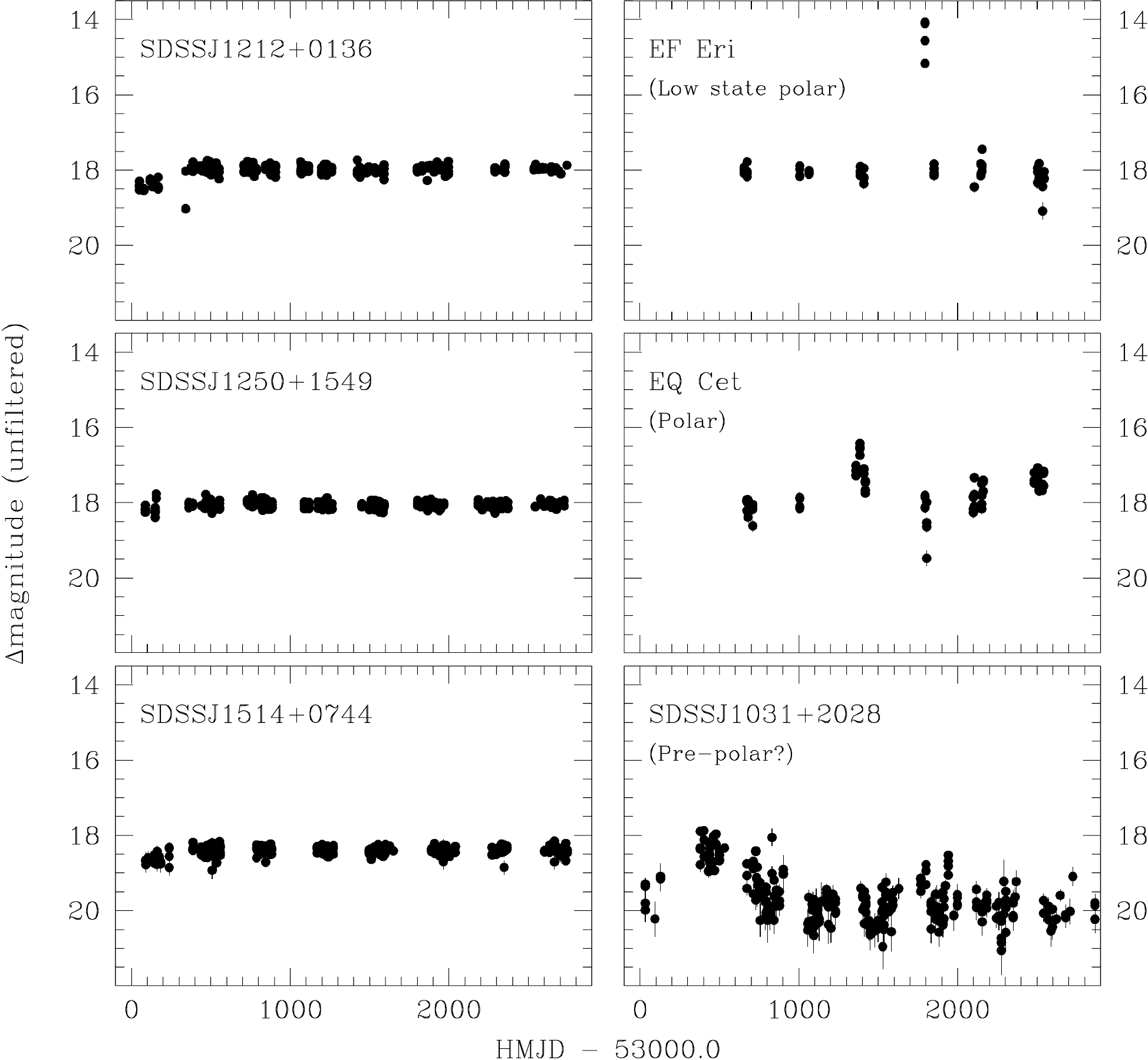}}
\caption{\label{fig:lcs} CRTS lightcurves of short period magnetic binaries, as labelled. 
SDSSJ1212+0136, SDSSJ1250+1549 and SDSSJ1514+0744 display no large scale 
variability, eclipses or high states. The lightcurve of EF~Eri shows a brief 
high state in 2008 November. The short period polar EQ~Cet displays large amplitude 
variability on timescales of years. SDSSJ1031+2028 is currently classified as a 
pre-polar. Its 42~MG magnetic field shifts the cyclotron harmonics to optical wavelengths, 
so the variability in the early part of the lightcurve could be due to variable cyclotron 
emission. The vertical scale is the same in all the panels to directly compare the 
variability amplitudes.}
\end{figure}

We also show the (unfiltered) lightcurves of SDSSJ1212+0136, SDSSJ1250+1549 and SDSSJ1514+0744, as obtained by the Catalina Real-time Transient Survey \citep[CRTS,][]{drake09crts}, in the left hand panels of Figure~\ref{fig:lcs}. The lightcurves span approximately three years, with four exposures per visit, taken over 30~minutes. The median separation between subsequent visits is 15 days for SDSSJ1250+1549 and 18 days for SDSSJ1514+0744.
We find no large scale variability, eclipses or high states of these binaries. Figure~\ref{fig:lcs} also includes the lightcurves of the low state polar EF~Eri, the pre-polar EQ~Cet and the possible pre-polar SDSSJ1031+2028, for comparison. These systems will be discussed in more detail in Section~\ref{sec:discuss}.

\section{Analysis and results}

\subsection{Orbital periods}

\begin{figure}
\centering
\includegraphics[width=6.5cm]{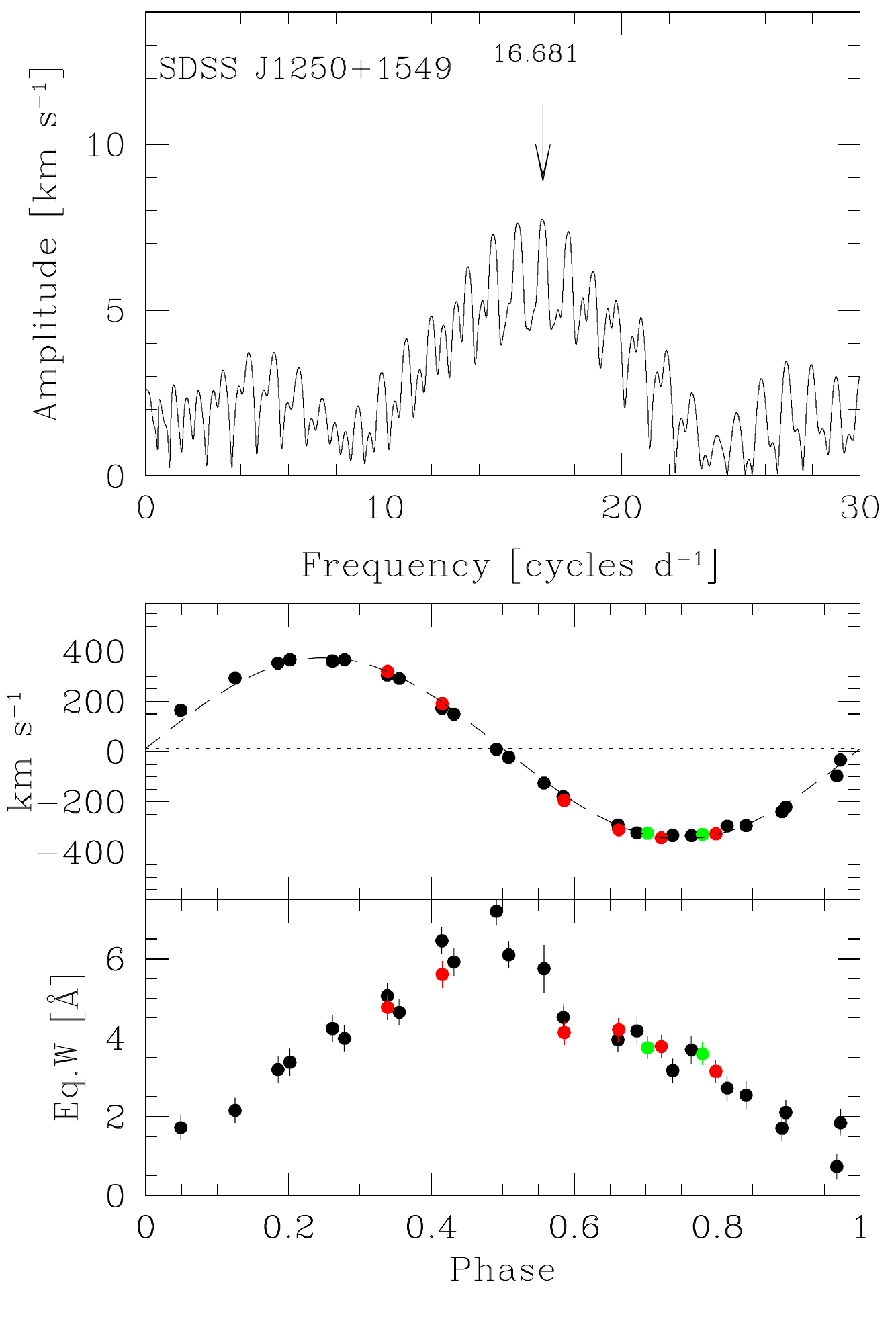}
\caption{{\em Top panel:} Amplitude spectrum calculated from the radial velocities 
of SDSSJ1250+1549. The adopted orbital frequency is indicated by an arrow. 
{\em Middle panel:} The measured \ha\, radial velocities folded on a period of 86.3
minutes. Data from the first night are plotted in black filled dots, the second
night in red and the third night in green. Error bars are plotted but in most
cases are smaller than the plot symbols. The dashed line is the best-fit
sinusoid to the radial velocities (Eqn.~\ref{eq:sine}). Its parameters are listed in
 Table~\ref{tab:prop}. 
{\em Bottom panel:} Equivalent width variation of the emission line, folded on 
the same ephemeris. 
\label{fig:periodJ1250}}
\end{figure}

\begin{figure}
\centering
\includegraphics[width=6.5cm]{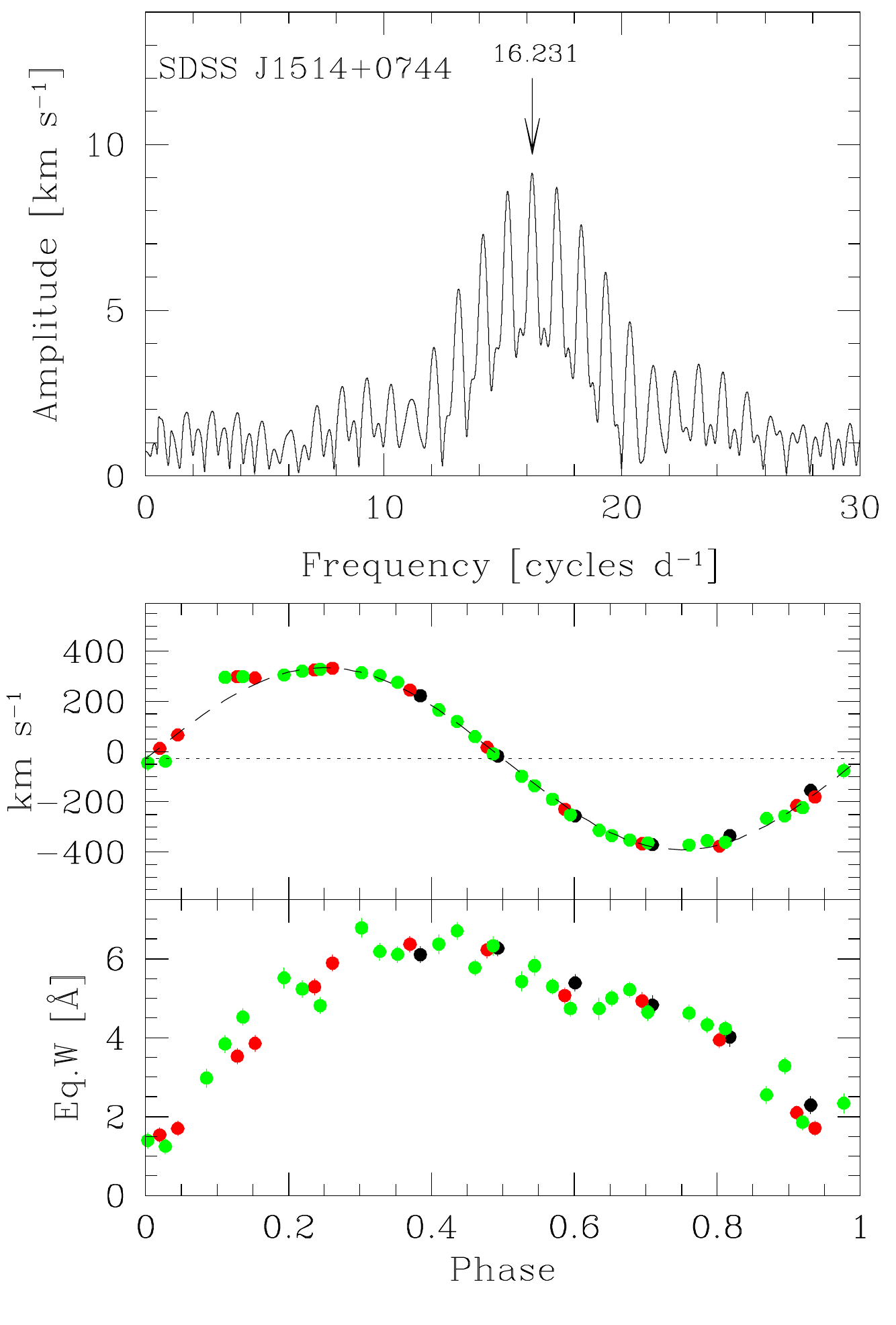}
\caption{As in Fig.~\ref{fig:periodJ1250}, but for SDSSJ1514+0744. The \ha\,
radial velocities  (middle panel) and equivalent width (bottom panel) are folded on 
the orbital period of 88.7 minutes.  Error bars are plotted but in most
cases are smaller than the plot symbols. \label{fig:periodJ1514}}
\end{figure}

\begin{figure}
\centering
\rotatebox{270}{\includegraphics[width=6cm]{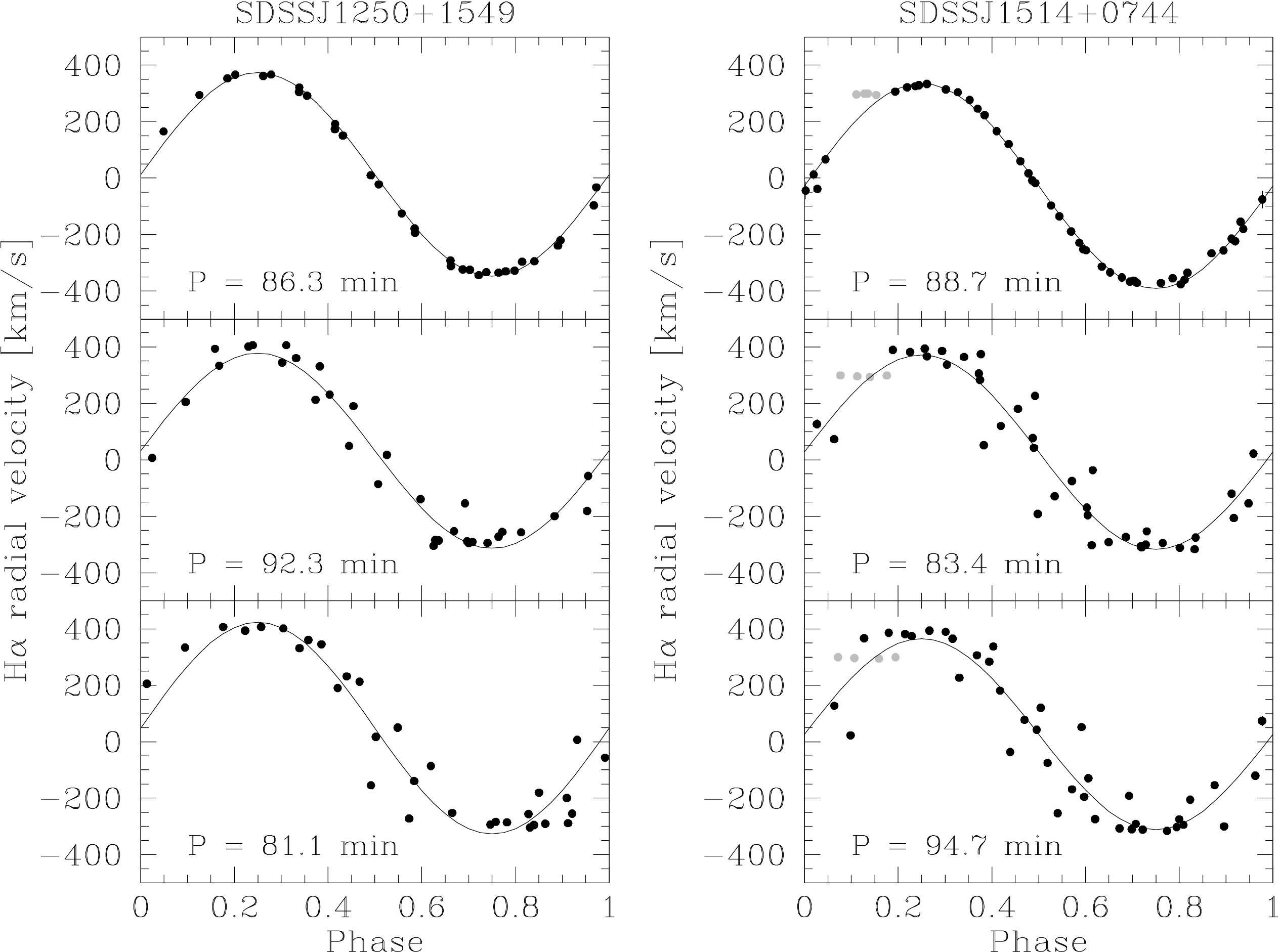}}
\caption{Radial velocities folded on competing one day aliases, as labelled. The strongest 
alias in each case (top panels) is well described by a sinusoidal fit, while folding the radial 
velocities on the nearest aliases lead to much larger scatter.  Error bars 
are plotted but are smaller than the plot symbols. Points in gray were not included in the fit.
 \label{fig:aliases}}
\end{figure}

We measured the radial velocities by fitting a single Gaussian function to the narrow \ha\, emission line in each spectrum. The local continuum was accounted for by including a pivoting straight line in the fit. 
We carried out a Lomb-Scargle \citep{lomb76,scargle82} and orthogonal polynomial \citep[ORT,][]{sc96_ort} analysis on our time series data to search for periodic signals. The amplitude spectra are shown in Figures~\ref{fig:periodJ1250} and \ref{fig:periodJ1514}, as these have the advantage that the variability at different frequencies can be compared directly in physical units. The strongest signals are found at 86.3 minutes (SDSSJ1250+1549) and 88.7 minutes (SDSSJ1514+0744), and are labelled on the plots. 
To confirm that these are the correct periods to associate with the orbital period of each system, we compare the folded radial velocity curves in Figure~\ref{fig:aliases}. A sinusoidal fit of the form 
\be V(t) = \gamma + K_2\;\sin\,[2\pi(t-\mbox{HJD}_0)/\Porb] \label{eq:sine} \ee
is overplotted on the measured radial velocities. HJD$_0$ is selected such that it corresponds to the red to blue crossing of the radial velocities. In both cases the strongest signal results in a radial velocity curve which is well described by a sinusoidal fit (Fig.~\ref{fig:aliases}, top panel), while folding the radial velocities on the one-day cycle count aliases (bottom two panels) leads to much larger scatter and $\chi^2$ values which are at least 15 times larger. We also carried out a bootstrap analysis to assess the importance of individual measurements in determining the orbital period. For each system, we calculated the periodogram from 10\,000 randomly selected subsets of the radial velocity curve, and recorded the frequency of the strongest peak in each case. For SDSSJ1514+0744, 99.95 per cent of the subsets returned 88.7 minutes as the strongest period, and in the case of SDSSJ1250+1549, 98.13 per cent selected the 86.3 minute alias as the strongest signal. 
We are therefore confident that the strongest peaks in Figures~\ref{fig:periodJ1250} and \ref{fig:periodJ1514} correspond to the respective orbital frequencies.

To illustrate the phase coverage of each night's observations, we plot the folded radial velocities in different colours in the middle panels of Figures~\ref{fig:periodJ1250} and \ref{fig:periodJ1514}. The parameters of the best-fit circular orbit (Eqn.~\ref{eq:sine}) and their uncertainties are shown in Table~\ref{tab:prop}. The deviation from the sine wave at phases $\phi\sim0.10-0.15$ of SDSSJ1514+0744 is real and due to an additional component to the \ha\, emission (see Section~\ref{sec:dopplermaps}). These higher radial velocities are seen at the same phase in subsequent orbits and on different nights, and were excluded from the sine fit.

The Zeeman components of the white dwarf's absorption lines are clearly variable as well, both in position and in strength. By eye, these variations do appear to track the orbital motion, but, due to the presence of the rotational magnetic field, which modulates the field strength over the visible hemisphere, the absorption lines are unreliable as an indicator of the white dwarf's radial velocity.

\begin{table}
   \caption{Properties of the new magnetic binaries. \label{tab:prop}}
   \begin{tabular}{l c c c}
      \hline
         ~                    & SDSSJ1250+1549     &   SDSSJ1514+0744     & Ref. \\
      \hline
      $\Porb$ (d)             & $0.05995(1)$       & $0.061610(1)$        &   1  \\
      HJD$_{0}$               & $2455646.7056(1)$  & $2455646.89093(3)$   &   1  \\
      $K_{\mbox{\small em}}$  & $360.89\pm2.54$    & $362.84\pm0.70$      &   1  \\
      $\gamma$ (km/s)         & $12.83\pm2.10$     & $-27.98\pm0.77$      &   1  \\
      \Twd (K)                & 10\,000            & 10\,000              &  2,3 \\
      $\log\:g\:$ (assumed)   & 8.0                & 8.0                  &   2  \\
      $B$ (MG)                & $20\pm2$           & $36\pm4$             &   2  \\
      $D$ (pc)                & $163\pm33$         & $211\pm42$           &   1  \\
   \hline
   \end{tabular}
   \begin{flushleft}   
      \small{$^1$ This paper}\\
      \small{$^2$ \citet{kulebi09}}\\ 
      \small{$^3$ \citet{vanlandingham05}}\\
    \end{flushleft}   
\end{table}

\subsection{White dwarf properties} \label{sec:wd}

SDSSJ1250+1549 and SDSSJ1514+0744 were both previously identified as magnetic white dwarfs from their SDSS spectra \citep{vanlandingham05, kulebi09}. Using a grid of magnetic white dwarf model spectra, along with magnetic dipole modelling, \citet{kulebi09} estimate their dipole magnetic field strengths to be 20~MG and 36~MG respectively. The white dwarf temperature was found to be 10\,000~K in both systems,
using a grid stepsize of 3000~K and assuming a canonical surface gravity of $\log\:g\:=8.0$. The same temperature was obtained by \citet{girven11}, by fitting non-magnetic white dwarf models to the
SDSS $ugri$ photometry. The best-fit models are shown in Figure~\ref{fig:seds}. We summarise all derived properties of the two new magnetic binaries in Table~\ref{tab:prop}.

For a 0.6\msun white dwarf which has never accreted, a temperature of 10\,000~K corresponds to a cooling age of 0.6~Gyr \citep{holbergbergeron06}\footnote{We used their updated grid of DA white dwarf cooling models, available at\\ http://www.astro.umontreal.ca/$\sim$bergeron/CoolingModels/.}. A more massive white dwarf of the same temperature will be older, increasing to 1~Gyr for 0.8\msun, or 2~Gyr for a 1\msun white dwarf. The surface layers of an accreting white dwarf will be kept hotter by compressional heating of the accretion activity \citep{townsleybildsten02}, so the cooling age sets a lower limit on the age of the binary.

We can use the white dwarf temperature and apparent magnitudes to estimate the distance to these binaries. Assuming a white dwarf radius of $0.01$\rsun we calculate the distance to SDSSJ1250+1549 to be $163\pm33$~pc and the distance to SDSSJ1514+0744 to be $211\pm42$~pc. The uncertainties on these distance estimates allow for a $\pm1000$~K variation of the white dwarf temperature \citep{girven11} and a $\pm0.25$ variation in $\log g$.

The \rosat\, non-detections places an upper limit of $\lesssim 6\times10^{-14}$ \ecs on the 0.1--2.4~keV X-ray 
flux, which for the assumed distances translate to an accretion rate lower than $5\times10^{-14}$\msun\,yr$^{-1}$. 
Note however that this estimate does not take the accretion luminosity emitted as cyclotron emission into account.

\subsection{Location of the \ha\, emission} \label{sec:location}

\subsubsection{\ha\, equivalent width} \label{sec:eqw}

The high radial velocity amplitude of the \ha\, emission line, as well as the fact that its strength is modulated on the orbital period, suggest that it is emitted from the surface of the companion star. To test this hypothesis, we measured the equivalent width of the emission line and folded it on the orbital ephemeris. To make an accurate
measurement, we had to account for radial velocity motion of the emission line through the absorption trough of the
white dwarf. We fitted the absorption line with the sum of a narrow and broad Gaussian, which was found to describe its shape well. The local continuum was normalised using the best fit double Gaussian for each spectrum before measuring the equivalent width of the emission line in the standard way. The results are shown in the bottom panels of Figures~\ref{fig:periodJ1250} and \ref{fig:periodJ1514}, folded on the orbital period.  
In both systems the equivalent width peaks at phase $\phi=0.5$. As defined by our spectroscopic ephemeris this phase corresponds to superior conjunction of the companion star, when we are viewing its irradiated side face-on. 
At first glance this type of modulation of the emission line strength suggests that the \ha\, emission is the result of irradiation by the white dwarf. However, \ha\, can also be produced by chromospheric activity of the companion star, and the two effects may be difficult to disentangle. During 2007 May and June \citet{howell08} found the \ha\, emission of SDSSJ1212+0136 essentially absent from the optical spectrum, over the full orbit. 
It had returned by 2008 February, but without a significant change in the overall flux level of the binary or the white dwarf dominated part of the spectrum. The authors therefore conclude that the \ha\, emission cannot be produced by irradiation. Interestingly, the same behaviour has been observed in EF~Eri as well \citep{howell06}. 
It is worth noting that irradiation by a 10\,000~K white dwarf will only raise the temperature of its companion's irradiated hemisphere by a few hundred Kelvin, for the orbital separations of these binaries, even assuming synchronised rotation and zero albedo. Using irradiated atmosphere models and observations of the polar VV~Pup ($\Porb=100$~min, \Twd=12\,000~K; \citealt{araujobetancor05b}) in its low state, \citet{mason08} show that irradiation cannot explain the observed line strengths for white dwarf temperatures below 20\,000~K, and that
most of the line emission must have its origin in stellar activity.  
If there is accretion taking place in the binary, even at a low level, the accretion spots on the white dwarf can easily reach the temperatures required to produce ionising photons and cause observable irradiation of the companion star, even if the white dwarf itself is too cool.  
The shape of the phase-folded equivalent width lightcurves of SDSSJ1250+1549 and SDSSJ1514+0744 does mean that if the \ha\, emission here is due to stellar activity, it must be restricted to, or greatly enhanced, on the hemisphere of the companion facing the white dwarf only.  Such a scenario could result from magnetic interaction between the white dwarf and its companion, which would induce activity in the companion preferentially on the white dwarf facing hemisphere \citep{howell06}.

\subsubsection{Doppler tomography}  \label{sec:dopplermaps}

Doppler tomography \citep{marshhorne88,marsh05} provides an alternative way to locate the source of the emission. The method uses the fact that the observed line profile at a given orbital phase is the result of the integrated emission at the line wavelength, moving at different velocities in different parts of the binary system. The binary rotation allows us to view the system from different angles and separate the different velocity components. Such a reconstructed ``velocity image'' can then be compared to the expected velocities of the binary components to infer the origin of the emission.

\begin{figure}
\flushleft
\includegraphics[width=9.0cm]{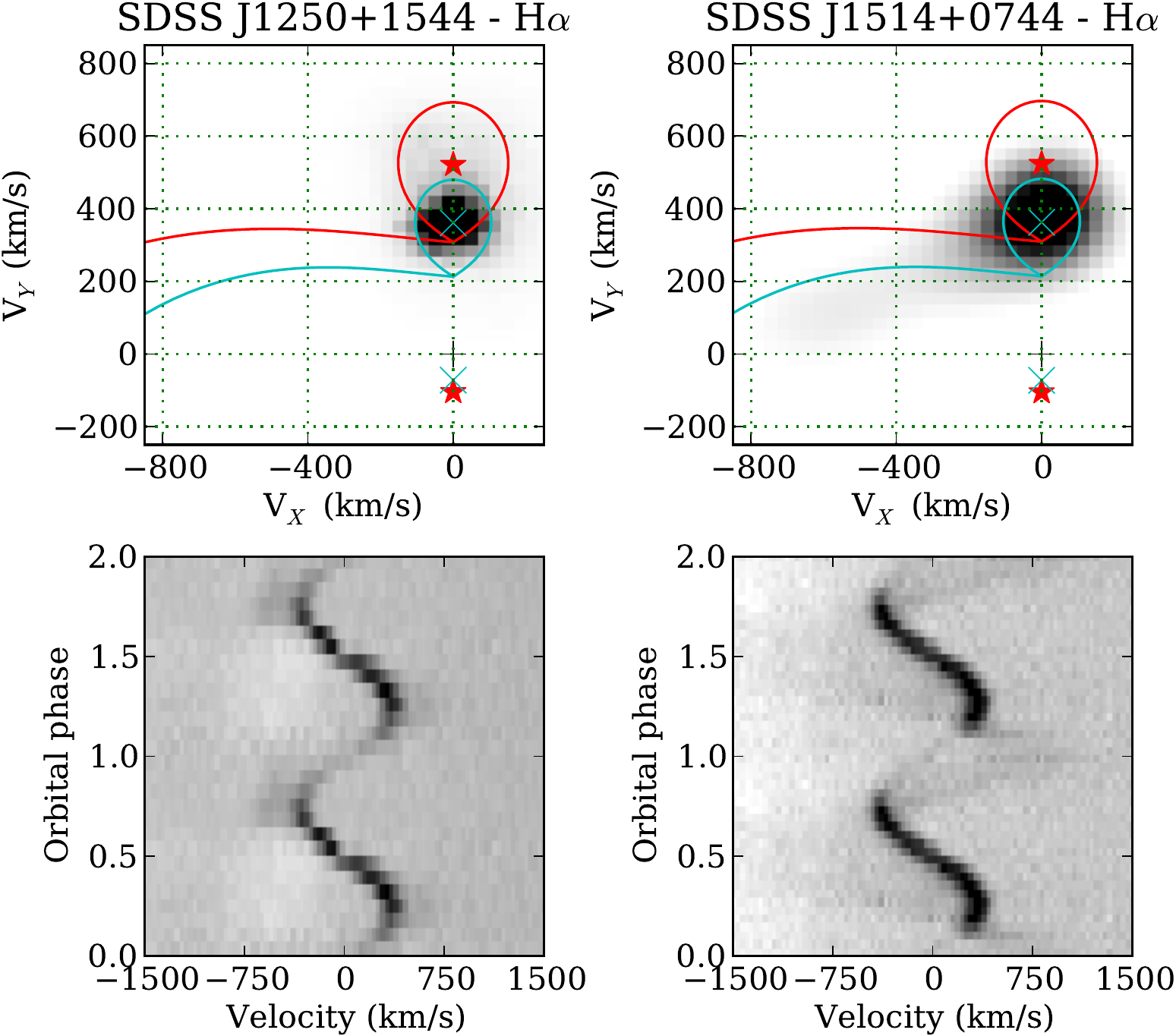}
\caption{Doppler tomography and trailed spectra of SDSSJ1250+1549 (left) and
SDSSJ1514+0744 (right). The Roche lobe of the secondary star in a binary with 
$q=0.2$ is overplotted on the maps as a reference, for two values of the 
$K$-correction. The smaller Roche lobe (centre of mass marked by a $\times$
symbol) has a zero $K$-correction, i.e. assumes that the \ha\, emission is
distributed uniformly over the surface of the star. The larger Roche lobe 
(star symbol) is calculated assuming the emission arise from the irradiated 
inner face of the secondary star only, with no contribution from the
cooler opposite hemisphere. The $+$ symbol at (0,0) indicates the centre of mass 
of the system. The predicted trajectory of a ballistic gas stream emerging from 
the inner Lagrangian point in each case is shown as well. The trailed spectra 
in the lower panels are phase binned, and two cycles are plotted for clarity. 
Note the faint S-wave of the mass transfer stream in the trail of SDSSJ1514+0744.
\label{fig:dopplermaps}}
\end{figure}

Our Doppler tomograms are shown in Figure~\ref{fig:dopplermaps}, along with the trailed spectra they were computed from. Here we simply subtracted polynomial fits to the continuum from the spectra without removing the white dwarf absorption or any other features of the spectrum. For comparison, Doppler maps are overplotted with the Roche lobe of a secondary star in a binary with mass ratio $q=M_2/M_1=0.2$, for an assumed minimum and maximum value of the $K$-correction (see Section~\ref{sec:Kcorr}). The corresponding velocity projections of a ballistic stream emanating from the inner Lagrangian point are also shown. 

The strong emission line dominating the trailed spectra maps to a single bright point in the tomogram, consistent with the \ha\, emission originating on the companion star. No other contribution is seen in the  tomogram of SDSSJ1250+1549 (left hand panel), but SDSSJ1514+0744 displays additionally some faint emission which appears to be active mass transfer between the two stars. The stream does not follow a ballistic path for any realistic set of
binary parameters. It has a lower velocity than the ballistic stream in all cases. This feature could be evidence for wind accretion by the white dwarf \citep{ww05}. Such an interpretation of the tomogram should be treated with some caution. A magnetically controlled outflow will have a velocity component out of the plane of the binary orbit, violating a key assumption of Doppler tomography. Nevertheless, this technique has been successfully applied to polars with magnetically channelled accretion flows in the past \citep[e.g.][]{schwope04_polardopp}, often showing a ballistic gas stream leaving the L1 point before it is disrupted and channelled by the magnetic field
within the Roche lobe of the white dwarf. The additional emission component is also visible as a faint S-wave in the
trailed spectra and as a radial velocity disturbance in Figure~\ref{fig:periodJ1514}.

\subsubsection{K-correction of the \ha\, line}  \label{sec:Kcorr}

We have shown in Section~\ref{sec:location} that the \ha\, emission originates mainly from the inner hemisphere of the companion star. Therefore, the measured emission line radial velocity amplitude ($\Kem$) corresponds to the velocity of the centre of light, rather than the centre of mass of the secondary star. As the centre of light is closer to the centre of mass of the binary, $\Kem$ underestimates the true centre of mass radial velocity, $K_2$.

In order to place a dynamical constraint on the white dwarf mass, a correction factor needs to be applied. This correction factor is a function of the mass ratio $q=M_2/M_1=K_1/K_2$ and the distribution of the \ha\, emission across the surface of the star. The latter is difficult to determine exactly, but it can be constrained through geometrical considerations \citep{wadehorne88,munosdarias05}. %
The corrected radial velocity amplitude of the secondary star is given by
\be K_2 = \frac{\Kem}{1-(1+q)fR_2/a} \label{eq:K2Kem}\ee %
where $fR_2$ is the distance of the centre of light from the centre of mass of the secondary star, with $0\leq f\leq 1$. $f=0$ represent the case where \ha\, is emitted uniformly across the surface of the secondary star, so that $K_2=\Kem$, while $f=1$ is the opposite extreme where all the emission originates in a small spot on the secondary star closest to the white dwarf. For the case where the hemisphere facing the white dwarf emits uniformly, and there is no contribution from the cool hemisphere, $f=4/3\pi\approx0.42$ \citep{wadehorne88}. We will assume this value as representative in the calculations that follow.
For two stars separated by a distance $a$, orbiting their common centre of mass with a period $\Porb$, the orbital velocity of the secondary star is given by %
\be K_2 = \frac{2\pi a\sin i}{\Porb(1+q)} \label{eq:K2Porb}\ee %
where $i$ is the binary inclination. Combining equations~\ref{eq:K2Kem} and \ref{eq:K2Porb} allows us to write the measured emission line velocity $\Kem$ as a function of the mass ratio $q$ and the binary inclination $i$,
\be \Kem = \frac{2\pi a\sin i}{\Porb(1+q)}\left[1-(1+q)\frac{fR_2}{a}\right]. \label{eq:Kem}\ee %
We return to this equation in Section~\ref{sec:evstatus}.

\subsection{Spectral type of the secondary star}

No features of the secondary star, apart from the \ha\, emission, are detected in our time series spectra.  The UV-optical part of the SED is well described by a $10\,000$~K white dwarf in both systems but there is a large excess in the UKIDSS bands (Figure~\ref{fig:seds}). \citet{steele11} modelled the excess in SDSSJ1250+1549 with an M8 companion, but were only able to match the $JH$ flux with this spectrum. The enhanced $K$ band flux is likely due to cyclotron emission, but its overall contribution to the spectrum cannot be determined from these data alone. In general, the cyclotron emission is strongly modulated on the orbital period, making the UKIDSS `snapshot'
observations unsuitable for SED modelling.

For SDSSJ1514+0744, the cyclotron harmonics will appear at shorter wavelengths than in SDSSJ1250+1549, due to its stronger magnetic field. Cyclotron emission will almost certainly contribute to some or all of the NIR flux measurements but by an (as yet) unknown amount.  We note however the very low $J$ band flux of SDSSJ1514+0744 (see
Figure~\ref{fig:seds}). There is no error flag on  this measurement in UKIDSS, so it appears to be real. If so, it offers a constraint on the spectral type of the companion. Assuming that the cyclotron humps fall outside of the $J$ band in this system, so that the emission in this band is from the companion star only, the measured flux is consistent with an L3 type star or later at the derived distance of 211~pc. NIR spectroscopy will be essential to model the cyclotron emission and verify this estimate, as well as to search for other spectral features of the companion stars of both systems.


\section{Discussion}   \label{sec:discuss}

\subsection{Distinguishing between low state polars and pre-polars}

The component stars in polars and pre-polars are thought to be the same, but in the latter, accretion occurs via magnetic siphoning of the secondary's stellar wind \citep{ww05} rather than through Roche lobe overflow. The inferred low accretion rates of the pre-polars \citep[$\dot{M}\sim10^{-14}-10^{-13}$\msun\,yr$^{-1}$,][]{schmidt07} are compatible with this model, although an accretion stream has been detected in at least one system,
SDSSJ204827.9+005008.9 \citep{kafka10_2048}.

\begin{figure}
\centering
\includegraphics[width=7.5cm]{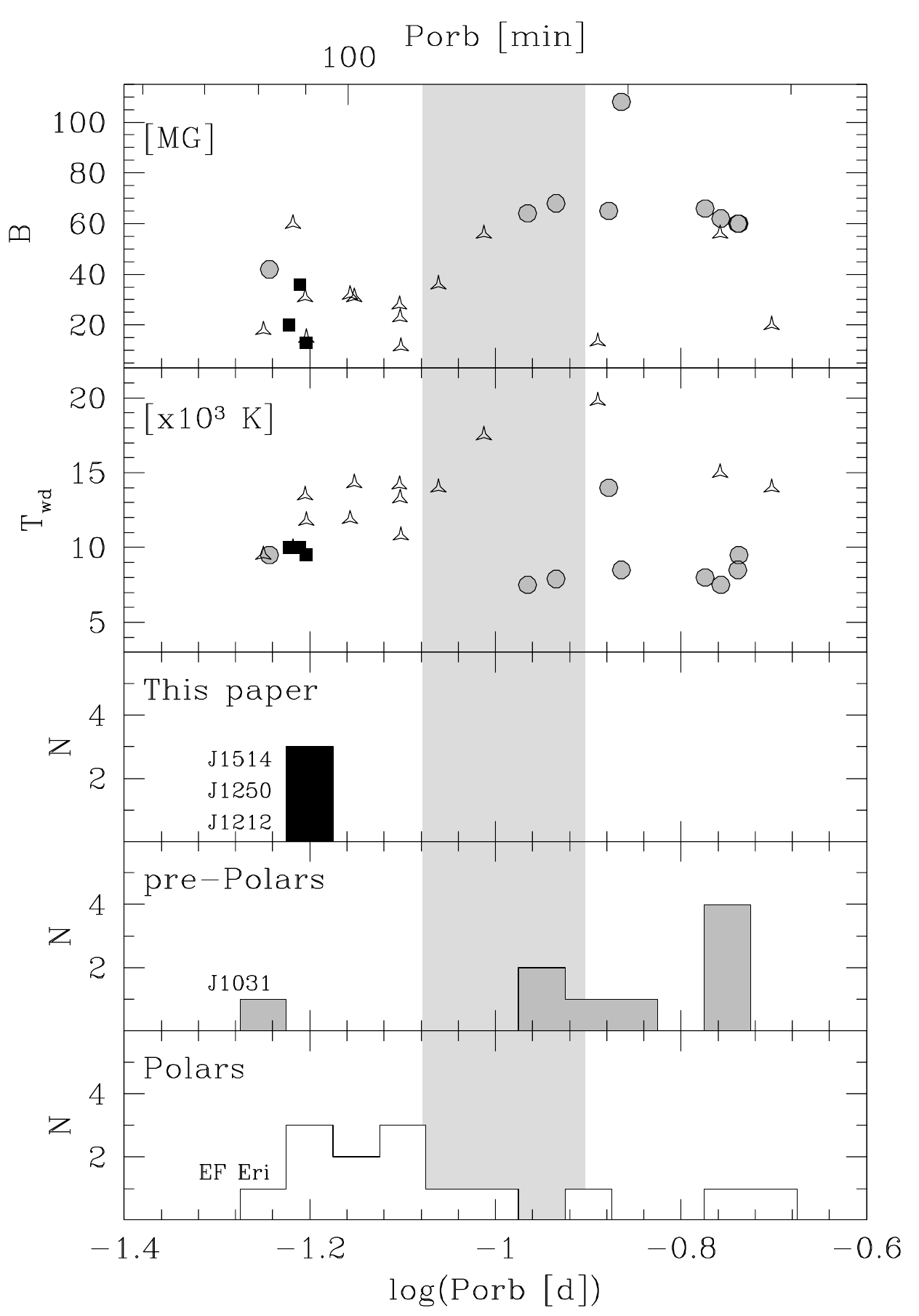}
\caption{\label{fig:allmagnetic} White dwarf magnetic field (top panel),
white dwarf temperature (second panel) and orbital period distribution 
(bottom three panels) of polars, pre-polars and the three magnetic binaries 
discussed in this paper. The light gray shading indicates the CV period gap. The 
`new' magnetic binaries (filled black squares, black histogram) have roughly
the same period and temperature and similar magnetic fields. The pre-polars
(gray dots, gray histogram) appear preferably at longer orbital periods,
slightly cooler white dwarf temperatures and higher magnetic fields. The polars
(open triangles, white histogram) span a wider range in temperature and
magnetic field than the other two classes. The parameter space 
\Twd$\leq10\,000$~K, $\Porb\leq90$~min, contains SDSSJ1212+0136, SDSSJ1250+1549, 
SDSSJ1514+0744, a known polar in a low state (EF~Eri), another polar previously observed
in both high and low states (FL~Cet) and a presumed pre-polar (SDSSJ1031+2028).}
\end{figure}

There are striking similarities between the two classes, such as the cool white dwarf, low inferred accretion rate and magnetic field strengths such that the cyclotron harmonics are shifted into the optical/NIR part of the spectrum. We compare graphically the properties of SDSSJ1212+0136, SDSSJ1250+1549 and SDSSJ1514+0744, with those of known polars and pre-polars in Figure~\ref{fig:allmagnetic}. We show only the polars for which a reliable
temperature estimate is available, taken from Table~1 of \citet{townsleygaensicke09}. 
As was already pointed out by \citet{schwope09}, the pre-polars are generally found at a slightly lower temperature than even the coolest polars. This can be understood in terms of accretion heating --- a higher accretion rate in polars leads to more compressional heating of the white dwarf \citep{townsleybildsten02,townsleygaensicke09}, raising the observed effective temperature. 

The orbital period is a potential discriminator between the two classes. The pre-polars all have periods longer than $2.5$~hours, with the exception of SDSSJ103100.6+202832.2 (shorthand notation: SDSSJ1031+2028) at 82~minutes
\citep{schmidt07,linnell10}. It seems plausible that the cooler, long period systems are still evolving towards Roche lobe contact, but it is important to continue monitoring these systems to reveal any high states which may prove them to be low state polars instead, as happened in the case of EQ~Cet \citep{schwope99,schwope02rosat}. Apart from being the only short-period pre-polar, SDSSJ1031+2028 is also the only one without evidence for a late type secondary star in its optical spectrum. With a magnetic field of 42~MG, the optical emission is dominated by cyclotron harmonics, and the spectrum has no clear features which can be attributed to the companion star. In fact, the $m=4$ harmonic is centred very close to \ha, complicating an analysis of the emission line variability. Based on the absence of a molecular band at 7600\AA (which falls between the cyclotron harmonics) \citet{schmidt07} are able to limit the spectral type of the secondary star in this binary to $\geq$M6, meaning that it is underfilling its Roche lobe for the 82 minute orbital period. We also note that its magnetic field is lower than the other pre-polars, but higher than that of EF~Eri, SDSSJ1212+0136 and the two new binaries discussed here. 

It is interesting to note the position of EF~Eri in the \Teff--$\Porb$ (second) panel of Figure~\ref{fig:allmagnetic}. It is the shortest period, lowest temperature polar on the plot, very close to SDSSJ1031+2028. 
EF~Eri was known as a typical high state polar from the time of its discovery \citep{williams79}, until it dropped
into a low state in 1997 \citep{wheatley_eferilow}. It has remained in a low state ever since, showing only brief bright states several years apart, but no return to its former high state\footnote{A long term monitoring lightcurve from the SMARTS telescope was presented by F.~Walter in 2009 March at the 14th North American Workshop on Cataclysmic Variables and Related Objects. The presentation available at http://www.noao.edu/meetings/wildstars2/talks/wednesday/walter.ppt}. 
One of these short bright states was detected by the CRTS in 2008 November and is shown in Figure~\ref{fig:lcs}. The lightcurve of EF~Eri shows no long timescale variability during its low state like the polar EQ~Cet in the panel below it. In fact, it displays greater similarity to the lightcurves of SDSSJ1212+0136, SDSSJ1250+1549 and SDSSJ1514+0744. The lightcurve of SDSSJ1031+2028 displays more variability, likely caused by the variable cyclotron emission dominating its optical spectrum. 

The general similarity between EF~Eri and SDSSJ1031+2028 was also noted by \citet{schmidt07} and it remains a possibility that SDSSJ1031+2028 could turn out to be a polar in an extended low state.

\subsection{Evolutionary status}  \label{sec:evstatus}

Where do the new binaries fit into this picture? The parameters of the three magnetic binaries, SDSSJ1212+0136, SDSSJ1250+1549 and SDSSJ1514+0744, are plotted in Figure~\ref{fig:allmagnetic} in filled black symbols. With their 10\,000~K white dwarfs, strong magnetic fields and orbital periods $80-90$ minutes, they occupy the same parameter space as EF~Eri and SDSSJ1031+2028. Apart from the variable \ha\, line, no spectral features of the companion star have so far been detected in their spectra. Flux considerations limit the spectral type of the companion star in SDSSJ1212+0136 to L5 -- L8 \citep{schmidt05b,farihi08}, but it is not clear yet whether this system is a detached binary or whether it is a polar currently in an extended low state.   
Our current data on SDSSJ1250+1549 and SDSSJ1514+0744 cannot constrain the parameters of these systems very tightly,
so we consider below three possible evolutionary scenarios: {\em i)} a polar in an extended low state, in which we assume that the secondary star fills its Roche lobe, {\em ii)} a detached binary which has not yet evolved to Roche lobe contact, and for which we assume a mass-radius relation for the companion star appropriate for isolated late type stars, and {\em iii)} a polar which has evolved past the orbital period minimum of CVs, and is now cooling and fading.

\subsubsection{Low state polar}  \label{sec:polars}

Under the assumption that these magnetic binaries are polars, we can take advantage of the fact that the secondary stars must be filling their Roche lobes, and approximate $R_2$ by the volume-equivalent Roche lobe radius, $R_L$. We use the expression derived by \citet{sirotkinkim09}, which models the secondary star as an $n=3/2$ polytrope,
\be \frac{R_L}{a} = \frac{0.5126\,q^{0.7388}}{0.6710\,q^{0.7349}+\ln(1+q^{0.3983})}\:. \label{eq:RL} \ee %
This approximation to the volume-equivalent radius offers an improvement over the \citet{eggleton83} relation 
for donor stars below the CV period gap which are fully convective and therefore less centrally concentrated. 

\begin{figure}
\centering
\includegraphics[width=7cm]{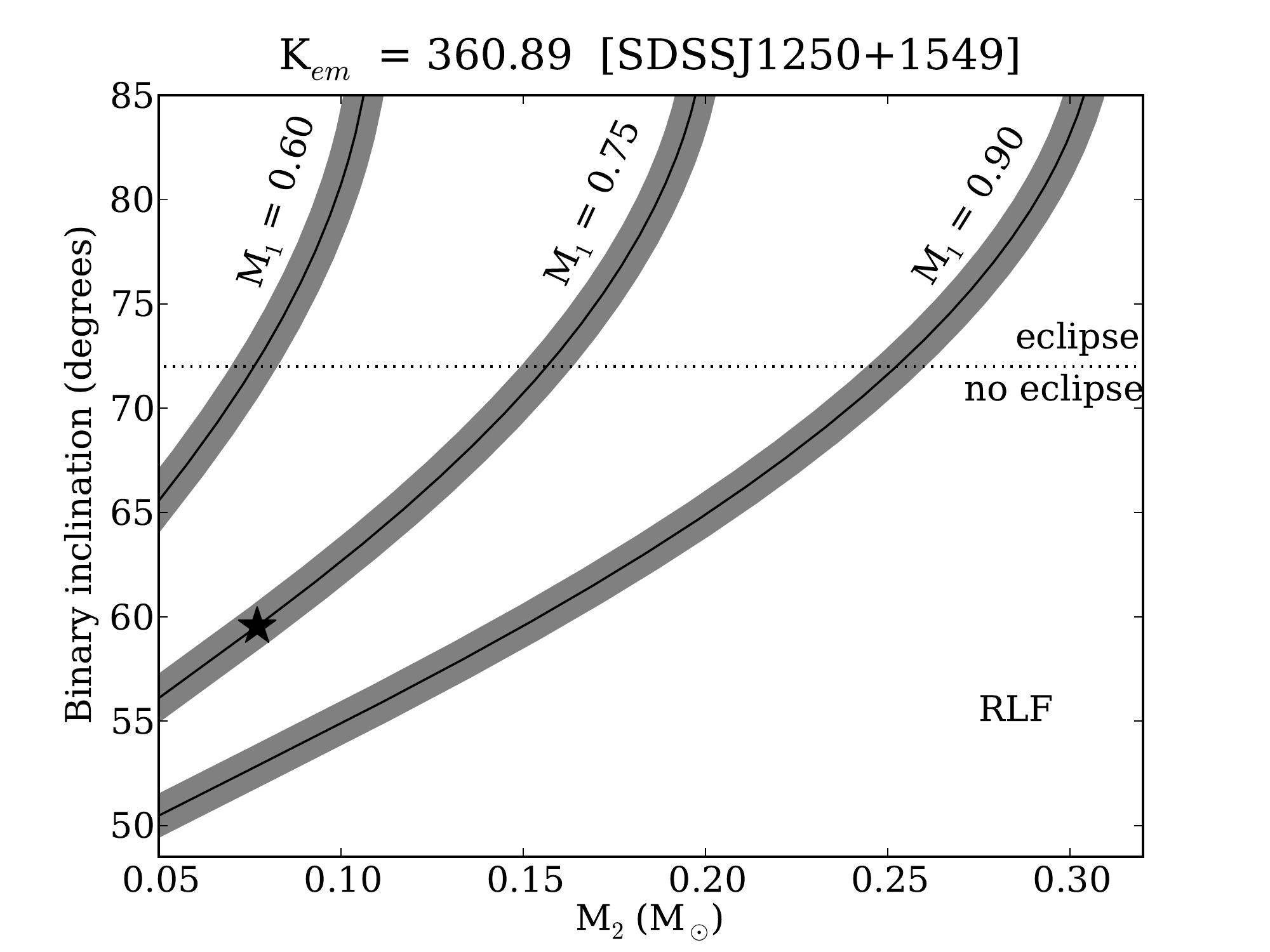}
\includegraphics[width=7cm]{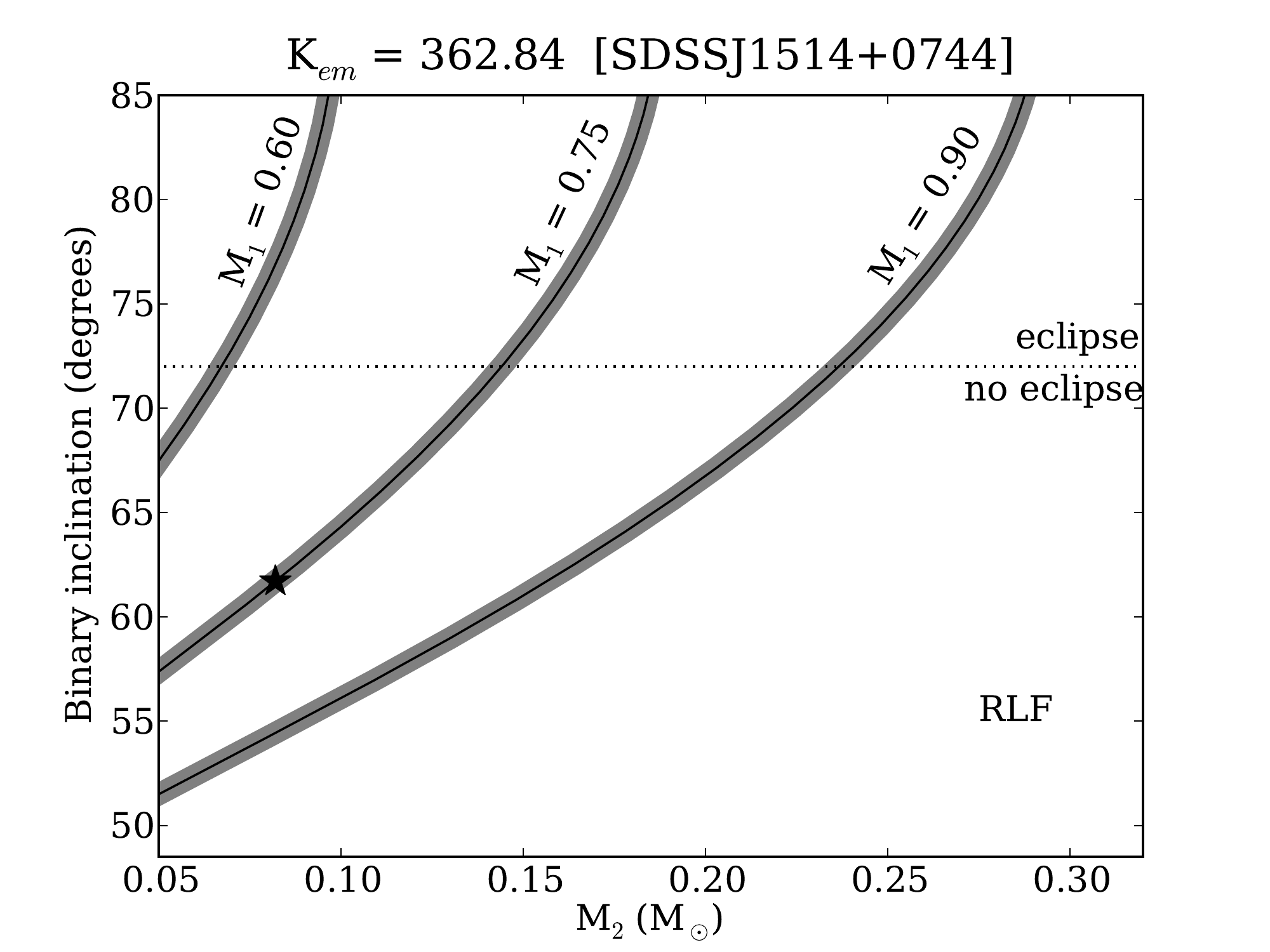}
\caption{\label{fig:Kem_RLF} 
The emission line velocity amplitude $\Kem$ is a function of both binary 
inclination $i$ and the mass ratio $q$ (see text for details).
The solid black lines are the $\Kem$ contours corresponding to the velocity 
amplitude measured from our spectra, for each pair of $M_2$,$i$ values, and 
the white dwarf mass $M_1$ as labelled. The corresponding error on $\Kem$ is 
indicated as gray shading. 
The area above the dotted horizontal line is ruled out as we find no evidence 
for an eclipse in our data. The binary parameters, as derived from the  
\citet{knigge11} donor star sequence, is indicated by a star symbol.
The `RLF' is a reference to the fact that this calculation assumes Roche lobe
filling.}
\end{figure}

Using equation~\ref{eq:Kem}, we can calculate all possible ($q$,$i$) combinations which give the measured emission line velocity $\Kem$. The results are shown in Figure~\ref{fig:Kem_RLF}. We fixed $M_1$ at 0.6\msun, 0.75\msun\, and 0.9\msun\, in three separate calculations, and connect the possible ($M_2$,$i$) pairs for each case with a black solid line. The binary inclination can be constrained based on the fact that we find no evidence for eclipses in our spectra or in the CRTS lightcurves. This restricts the inclination to $i<72\degr$. The inclination constraint of course depends on the on the mass ratio and on the radius of the secondary star. The limit we plot in Figure~\ref{fig:Kem_RLF} corresponds to the lowest inclination for the range of masses considered for which no eclipse of the secondary star would be observed. For the lowest values of $M_2$ eclipses would not be observed for inclinations up to $77\degr$. The lower limit on the inclination is set by the Chandrasekhar mass limit for
the white dwarf, restricting it to $i>38\degr$ in SDSSJ1250+1549 and $i>39\degr$ in SDSSJ1514+0744.

The CV donor star sequence of \cite{knigge11} is a semi-empirical sequence describing the physical and photometric properties of the secondary stars in cataclysmic variables, as a function of orbital period.  It was derived using non-magnetic CVs only, so strictly it only applies in the case of non-magnetic systems, but we include it as it may provide some insight into the properties of these binaries. The effect of the white dwarf magnetic field is to reduce the efficiency of magnetic braking of the secondary star by partially or fully controlling the mass flow \citep{wickramasinghewu94}. The reduced angular momentum loss results in a reduced mass transfer rate and therefore donor stars which are less out of thermal equilibrium than their non-magnetic counterparts at the same $\Porb$.

For the measured orbital periods of 86 and 89 minutes for SDSSJ1250+1549 and SDSSJ1514+0744 respectively, the typical non-magnetic CV donor has a mass of 0.077\msun and 0.082\msun, with radii 0.124\rsun and 0.129\rsun. The calculation assumes a representative white dwarf mass of $M_1=0.75$\msun, so we indicate these solutions as single points in Figure~\ref{fig:Kem_RLF}. The corresponding binary inclination values are $60\degr$ and $62\degr$.  

As a result of the lower mass transfer rate, the white dwarfs in polars are generally found to be cooler than non-magnetic CV white dwarfs \citep{sion04}. A 10\,000~K white dwarf temperature implies an average accretion rate 
over the thermal timescale of the envelope ($\sim10^5$ year) of $\dot{M}=1.2\times10^{-11}$\msun yr$^{-1}$ for a 0.9\msun white dwarf, or $\dot{M}=4.8\times10^{-11}$\msun yr$^{-1}$ for a 0.6\msun white dwarf \citep{townsleygaensicke09}.  These estimates are orders of magnitude higher than what we infer from the \rosat\, non-detections, but not atypical of the secular accretion rate of polars \citep[][p.347]{warnerbook}. The implication is that, unless they are younger than 0.6~Gyr, the white dwarf in these binaries must have accreted at a much higher rate in the relatively recent past.

\subsubsection{Detached binaries} \label{sec:detached}

In this section we consider these binaries under the assumption that they are detached pre-polars. In this case, the
secondary stars have never filled their Roche lobes and the only accretion that has taken place would be through 
magnetic capture of the secondary's stellar wind. Wind accretion proceeds at a much lower rate \citep[$\sim6\times10^{-14}$\msun\,yr$^{-1}$,][]{ww05} than typically seen in systems with Roche lobe overflow, so causes very little accretion-induced heating of the white dwarf. Unlike for CVs, the white dwarf temperature can therefore be used as an approximate age indicator. For a 0.6\msun\, white dwarf (resp. 0.9\msun) a temperature of 10\,000~K corresponds to a cooling age of 0.6~Gyr (resp. 1.3~Gyr) \citep{holbergbergeron06}$^{\mbox{\footnotesize 3}}$.

\begin{figure}
\centering
\includegraphics[width=7cm]{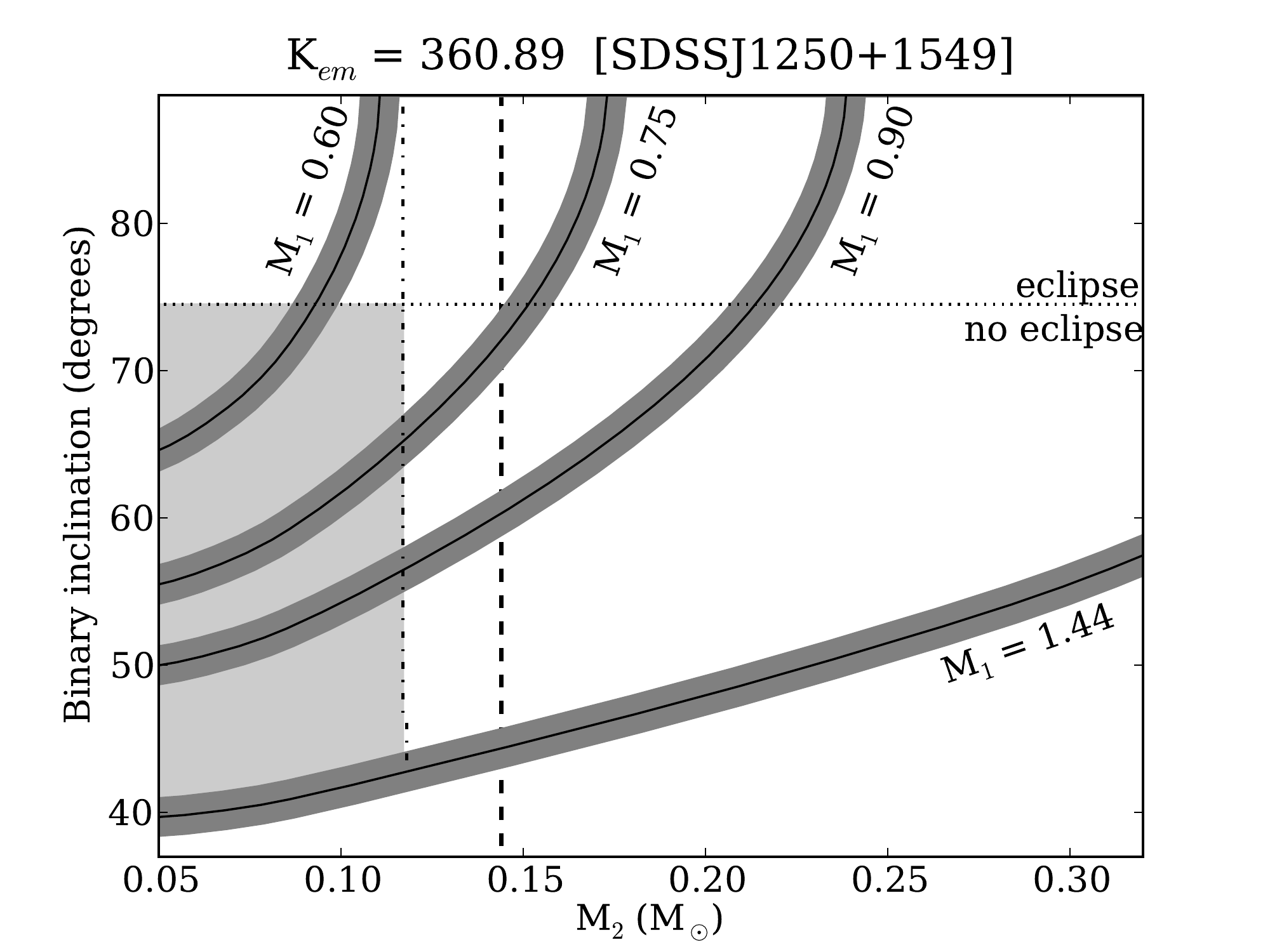}
\includegraphics[width=7cm]{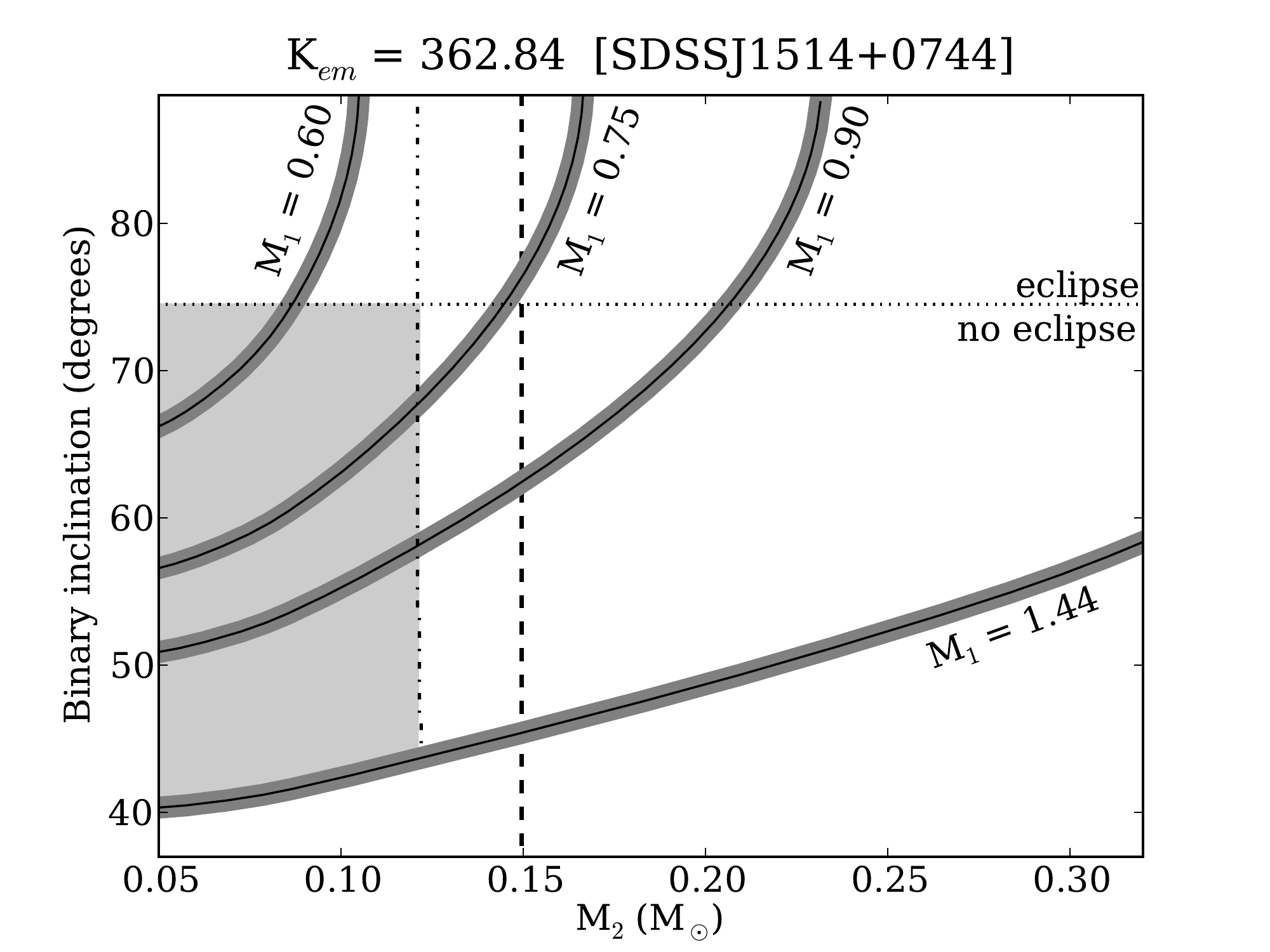}
\caption{\label{fig:Kem_det}  As for Fig.~\ref{fig:Kem_RLF}, but here we use the
mass-radius relation of an isolated M-dwarf of age 0.6~Gyr, rather than assuming 
a Roche lobe filling companion star. The thick dashed vertical line is a strict 
upper limit on the mass of the companion, from the density of the companion star. 
Requiring that the system remains detached, sets further limits on the mass (see text), 
leaving only the light gray shaded area as possible parameters.}
\end{figure}

\begin{figure}
\centering
\rotatebox{270}{\includegraphics[width=6cm]{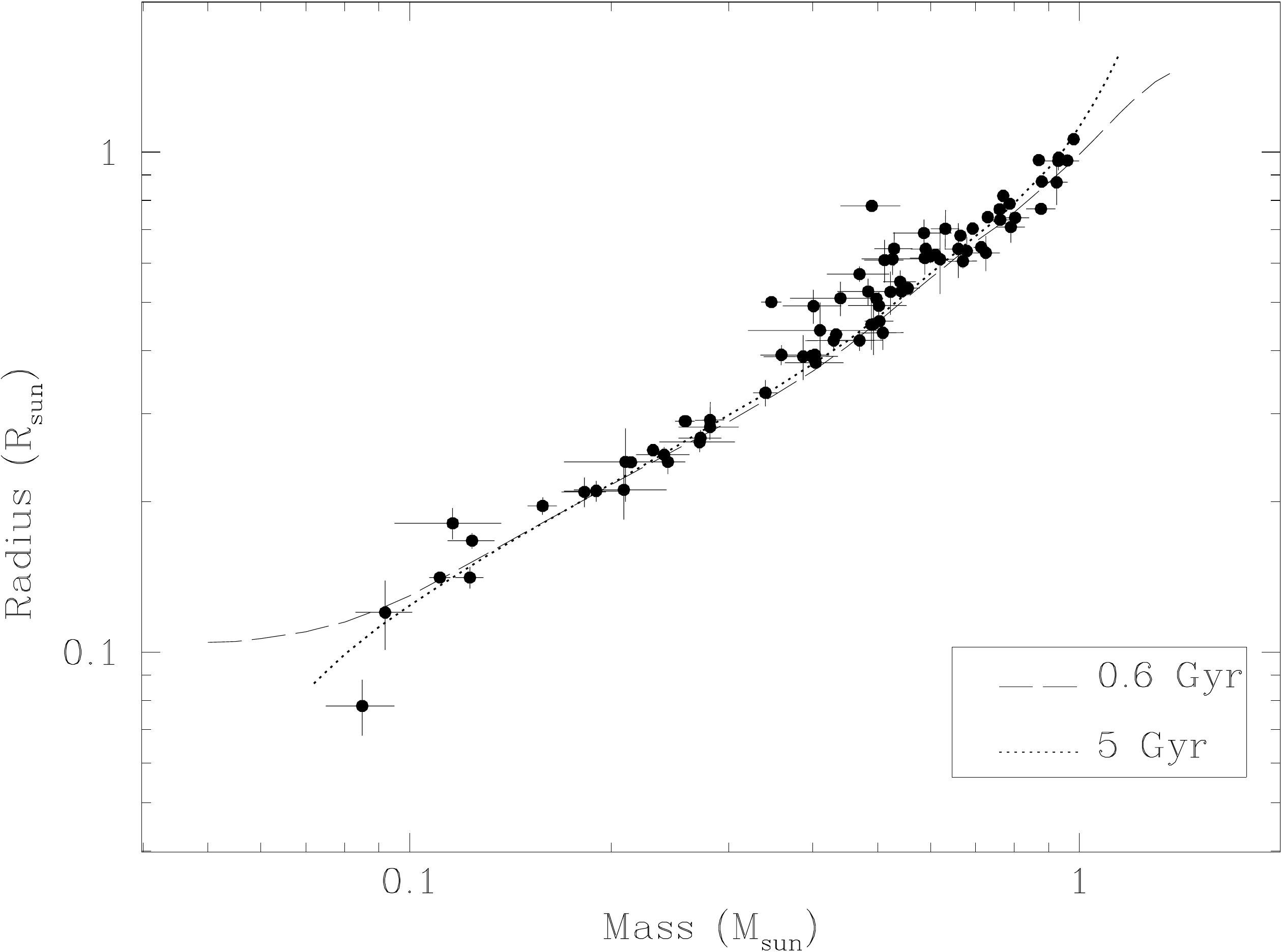}}
\caption{\label{fig:nointeract} The mass-radius relation of non-interacting
low-mass stars. The points with errorbars are measurements from literature, 
compiled by \citet{knigge11}. The dashed line is the 0.6~Gyr isochrone from 
\citet{baraffe98}, which is a lower limit on the age of the binaries considered 
here. The dotted line is for 5~Gyr. Note the deviation of the two models at low 
masses.}
\end{figure}

\begin{figure*}
\centering
\rotatebox{270}{\includegraphics[width=10cm]{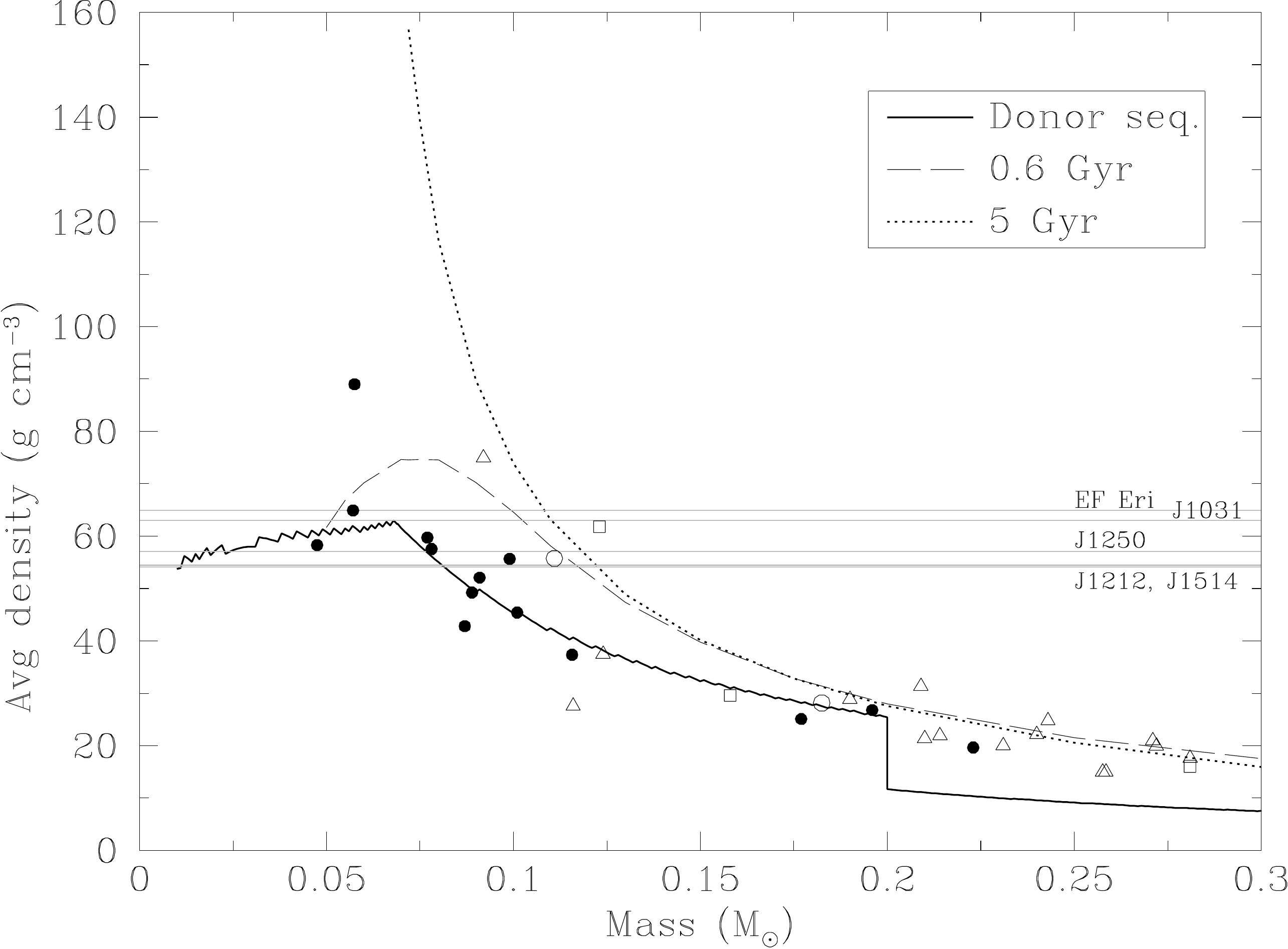}}
\caption{\label{fig:density} Average stellar density for different models of the
companion star. The dashed and dotted lines are the 0.6 and 5~Gyr low mass stellar 
models from \citet{baraffe98} and the solid black line the CV donor sequence from 
\citet{knigge11}. Overplotted as filled black circles are the average densities of
CV donor stars with a reliable mass and radius measurement from eclipse modelling 
\citep{savoury11}. The open circles are the companion stars in detached white dwarf 
binaries, the triangles represent low mass stars in other types of binaries 
and the squares are the measurements of single stars. The data were taken 
from the compilation by \citet{knigge11}.   A lower limit on the densities of
SDSSJ1031+2028, SDSSJ1212+0136, SDSSJ1250+1549 and SDSSJ1514+0744,
calculated from the CV period-density relation, are indicated by thin gray
lines. The density of EF~Eri is also shown, and as a known polar, its density must 
be close to the value shown.}
\end{figure*}

Due to this low mass loss rate, the structure of the companion stars is probably closer to that of non-interacting low mass stars, than to that of the companion stars of CVs. In Figure~\ref{fig:Kem_det} we show the range in inclination and companion star mass which can result in the observed emission line velocity $\Kem$. Rather than
assuming a Roche lobe filling secondary star as in the previous section, we use here the evolutionary models of \citet{baraffe98}, for low mass stars of ages 0.6~Gyr and 5~Gyr to determine the mass and radius of the companion stars. For simplicity, we show only the results for 0.6~Gyr in Figure~\ref{fig:Kem_det}. The 5~Gyr model results in a very similar diagram and does not affect the range of possible companion star masses derived below. The mass-radius relation for both models are shown in Figure~\ref{fig:nointeract}, along with measurements of non-interacting low-mass stars, as compiled from literature by \citet{knigge11}. Although the two models are very similar over a large part of the mass range under interest, they diverge slightly for masses below $M_2<0.15$, with the younger stars having a larger radius for a given mass. This slight difference in radius translates to a large difference in the mean stellar density for low mass stars. The mean density is proportional to $M/R^3$, so for a given (low) mass, the older stars are smaller, and therefore much denser. This difference is illustrated in 
Figure~\ref{fig:density}. For comparison, we also show the average density of the \citet{knigge11} CV donor star sequence, as well as a lower limit on the densities of the binaries considered in this paper. These limits, as indicated, were calculated from the period-density relation, which assumes that the stars are Roche lobe filling. 
If they are instead detached, as we assume in this section, their radii will be smaller for the same period and the densities higher. For non-detached binaries, the period-density relation therefore provides a lower limit on the mean stellar density. 

Since the orbital periods are well determined, we can also use the period-density relation to set an upper limit on the mass of the companion star. In Figure~\ref{fig:Kem_det}, the thick vertical dashed line indicates the mass of the companion star which have the same density as a Roche lobe filling star at the orbital period measured for these binaries. As before, we have used the mass-radius relation from \citet{baraffe98}.
It is clear from Figure~\ref{fig:density} that, for the same mass, the density of non-interacting stars (dashed and dotted lines) is higher than the CV donor stars (solid line). This is a result of the mass transfer which drives the donor stars of CVs out of thermal equilibrium, causing them to expand slightly. From eclipse measurements, \citet{littlefair08} find the donor stars in CVs to be $\sim10\%$ larger than isolated M-dwarfs of equal mass. 

The mass limit derived this way is a strict upper limit. If the companion stars were more massive, their radii would be larger (Figure~\ref{fig:nointeract}) and hence their densities lower (Figure~\ref{fig:density}). A larger radius means that the star would exceed the size of its Roche lobe so the binary would no longer be detached. We can constrain the mass further by requiring that, for each mass-radius pair, the binary remains detached, i.e. that the companion star radius is smaller than the Roche lobe radius. This is only true for $0.04<M_2/$\msun$<0.12$ (in both systems), shown in light gray shading in Figure~\ref{fig:Kem_det}. 

The mass range derived above allows for both a stellar or a substellar detached companion to the white dwarf. But how common are such binaries?   White dwarf plus M-dwarf binaries are plentiful, and a common product of the common
envelope evolution phase of binary stars. The SDSS DR7 catalogue of white dwarf--main sequence binaries, for example, contains 2248 entries \citep{rebassa11}, of which approximately one third are post common envelope binaries \citep{schreiber10}. However none of these binaries have a magnetic white dwarf like the binaries considered here. Brown dwarf companions to white dwarfs are also a rare find. From 4636 spectroscopically 
confirmed white dwarfs in the SDSS DR7, \citet{girven11} find less than 2 per cent to have an infrared excess consistent with a brown dwarf companion. In an earlier survey, \citet{farihi05} find fewer than 0.5 per cent of white dwarfs to have brown dwarf companions. Only seven white dwarf--brown dwarf binary systems have been spectroscopically confirmed, and only two of these are close binaries, GD\,1400 \citep{farihichristopher04} and 
WD\,0137-349 \citep{maxted06}.

Although we cannot rule out that SDSSJ1212+0136, SDSSJ1250+1549 and SDSSJ1514+0744 are detached systems, it seems rather unlikely that we would serendipitously find three near-identical detached magnetic binaries, all at a similar distance, in only $\sim2700$ degrees of sky (see Section~\ref{sec:spacedensity}). The implied space density of such a find is much higher than suggested by the small number of known pre-polars or non-magnetic white dwarf 
plus brown dwarf binaries.

The constraints derived from observable parameters leave more scope for these binaries to be Roche lobe filling (or nearly so) than detached. We note however that if these three binaries are pre-polars, it will extend (to lower values) the range of magnetic field strength and orbital period pre-polars are observed to have.

\subsubsection{Post-bounce polar}

In the standard model of CV evolution a system evolves to shorter periods through angular momentum loss, until a minimum period $\Pmin$, when the companion star can no longer sustain hydrogen burning in the core and becomes degenerate. The current observational estimate of this minimum period is $\Pmin=82.4$~minutes \citep{gaensicke09}. 
Further mass loss from the degenerate companion causes the binary orbit to expand and the system to evolve back to longer periods. It is observationally challenging to directly detect such a degenerate secondary in a CV, especially
since the optical luminosity is usually dominated by an accretion disc. The absence of an accretion disc in polars greatly improves chances of a direct detection, but even so it remains an elusive result. The third evolutionary state we consider for these binaries is that they are polars which have evolved past the period minimum.  

If we assume that the CV evolutionary track of \citet{knigge11} applies to SDSSJ1250+1549 and SDSSJ1514+0744, their orbital periods imply a binary age of $\gtrsim3.3$~Gyr. In the case of polars, the efficiency of magnetic braking is reduced by the magnetic interaction between the two component stars, so the evolution probably proceeds more slowly, making the binaries even older than this estimate. Nevertheless, even at $3$~Gyr, the spectral type of the donor is expected to be as late as T, with a mass of only $M_2\sim0.04$\msun. The mass transfer rate from the donor drops as the system evolves past $\Pmin$, so the low current accretion rate may be expected.

It is interesting to note that for the non-magnetic CVs of the \citet{knigge11} evolutionary track, the white dwarf temperature of a post-bounce CV at these orbital periods, is expected to be in the range 9160--11\,170~K. Although the observed temperatures of our magnetic binaries fall precisely in this range, it is not clear that
this estimate will apply to magnetic CVs. The reduced angular momentum loss rate in magnetic CVs results in a lower accretion rate and less compressional heating of the white dwarf, so \Twd\, is expected to be lower. By the same token, the evolution of magnetic CVs proceeds more slowly, so a magnetic CV is likely to be older and hence cooler than a non-magnetic CV at the same orbital period. An isolated 0.6\msun white dwarf will have cooled to 5500~K in 3~Gyr (7000~K for a 0.9\msun white dwarf), so the observed 10\,000~K must be result of relatively recent accretion at a high rate.

\subsection{Space density}  \label{sec:spacedensity}

The overlap in the SDSS DR7 and UKIDSS/LAS DR8 footprints is $\sim2700$ degrees \citep{girven11}. Our crossmatch revealed three magnetic white dwarfs with a low mass companion within this area, suggesting that the space density of such systems may be quite high. 
We estimate the effective survey volume searched by \citet{girven11} by integrating the volume enclosed by a spherical cone, taking into account the exponential drop-off of the space density perpendicular to the galactic disc (for more details, see Sect. 5.1 of \citealt{gaensicke09}).  We weigh the contributions of individual galactic latitudes by the fractional area covered by UKIDSS/LAS DR8. Adopting $d=200$~pc as a typical distance to the magnetic binaries discussed in this paper, and a scale height of either $H_z=190$~pc \citep{patterson84} or $H_z=260$~pc \citep{pretorius07b}, we find an effective survey volume of  $5.9\times10^5\mathrm{pc}^{3}$ and
$6.6\times10^5\mathrm{pc}^{3}$. 
Because of their intrinsic faintness, all three systems are within one scale height, i.e. the uncertainty in $H_z$ does not significantly affect our estimates, and we find  $\rho\simeq5\times10^{-6}\mathrm{pc}^{-3}$.  

This value is substantially higher than the estimated space density of ``genuine'' polars, $\simeq1.3-1.6\times10^{-6}\mathrm{pc}^{-3}$ \citep{thomasbeuermann98, araujobetancor05b}, and, in fact, comparable to the space density of non-magnetic CVs, $4{+6\atop-2}\mathrm{pc}^{-3}$ as estimated by \citet{pretorius07b} and \citet{pretoriusknigge12} from an analysis of two \rosat\, X-ray surveys. 

An alternative approach of assessing the relative frequency of the low-accretion rate magnetic binaries found by the SDSS, is to compare their number to that of equally bright non-magnetic SDSS CVs whose optical spectrum is dominated by their accreting white dwarfs. Within the Legacy footprint of SDSS DR7 \citep[8032~deg$^2$,][]{abazajian09_sdssdr7}, \citet{szkody11} have identified 15 CVs with $g\le18.84$ that clearly exhibit broad Balmer absorption lines from the white dwarf photosphere. This gives a surface density of $1.9\times10^{-3}\mathrm{deg}^{-2}$, which is comparable to the $1.1\times10^{-3}\mathrm{deg}^{-2}$ calculated for the low-accretion rate magnetic binaries. The spectroscopic completeness of SDSS is similar for both types of objects \citep{girven11}, and because of their clear observational hallmarks (non-magnetic CVs: Balmer emission lines from the accretion disc; low-accretion rate magnetic binaries: infrared excess due to cyclotron emission), they have been identified
from their SDSS spectra and other obeservations to a high degree of completeness as well.

Unless the sky distribution of the three low-accretion rate magnetic binaries discussed in this paper is subject to small-number anomalies, we have to conclude that they represent an abundant sub-population of close white dwarf binaries whose existence is not readily explained within the current theoretical framework.


\section{Summary and future observations}

We have presented VLT/FORS2 spectroscopy of two magnetic white dwarfs, SDSSJ1250+1549 and SDSSJ1514+0744 and have shown that they are binary systems, hosting low mass companion stars. We measured their orbital periods from a variable \ha\, emission line in their spectra and showed that in both systems this emission originates from the surface of the companion star. Their orbital periods are 86.3 and 88.7 minutes respectively. No other features that can be attributed to the companion star are seen in our spectra.
SDSSJ1514+0744 additionally shows evidence for a high velocity \ha\, component. Its velocity is lower than that of a ballistic stream, and may be interpreted as evidence of magnetic siphoning of the companion's stellar wind.
This strongly suggests that there is ongoing accretion in this system. In both binaries, the \ha\, equivalent width is also modulated on the orbital period, and peaks when the companion star is at superior conjunction. The \ha\, emission must therefore originate mainly from the hemisphere of the companion facing the white dwarf, although
it is not yet clear whether it is the result of irradiation by a hot spot on the white dwarf or magnetic activity on the companion star. The large radial velocity amplitude of the \ha\, line ($>360$~km\,s$^{-1}$) points to a high inclination in both systems, but we find no evidence of eclipses in the CRTS optical lightcurves of these binaries. 

Our current data cannot constrain the parameters of the binary components very tightly, so we have considered in detail three possible evolutionary scenarios: polar in an extended low state, detached pre-polar, or polar that has evolved past the orbital period minimum of cataclysmic variables. Although none of these possibilities can be ruled out, we favour the low state polar interpretation for the observed similarities between these binaries and EF~Eri, 
a well known polar which have been in a low state since 1997.

The properties of these two binaries are also near-identical to SDSSJ1212+0136, a known magnetic white dwarf plus possible brown dwarf binary. Incidentally, SDSSJ1212+0136 was also recovered by the method used to find these two new binaries. Just like  SDSSJ1212+0136, the SEDs of SDSSJ1250+1549 and SDSSJ1514+0744 display a large infrared excess, which  we have interpreted as cyclotron emission from the magnetic field of the white dwarf and possibly emission from a late type companion star.

Clearly these binaries will benefit greatly from further observations. In particular, an X-ray detection will give an improved estimate of the accretion rate, and time-resolved infrared spectroscopy will allow us to model the cyclotron emission, which will also constrain the accretion rate. Infrared spectroscopy will also allow us to look 
for spectroscopic signatures of the companion star, or place limits on its spectral type from flux considerations. Finally, an improved distance estimate will allow us to better constrain the white dwarf properties, which will help to discriminate between stellar and substellar models of the companion.



\section*{Acknowledgements}
EB, BTG, TRM and CMC acknowledge support from the UK STFC in the form of a Rolling Grant.
This paper is based on observations made with the European Southern Observatory's Very Large Telescope, as part of programme 086.D-0243B.
We also acknowledge use of the Sloan Digital Sky survey (SDSS, http://www.sdss.org), and the United Kingdom InfraRed Telescope (UKIRT) Infrared Deep Sky Survey (UKIDSS, http://www.ukidss.org). The SDSS is funded 
by the Alfred P. Sloan Foundation, the Participating Institutions, the National Science Foundation, the U.S. Department of Energy, the National Aeronautics and Space Administration, the Japanese Monbukagakusho, the Max Planck Society, and the Higher Education Funding Council for England.

\bibliographystyle{mn2e}
\bibliography{library2}  

\bsp

\label{lastpage}

\end{document}